# Toward a consistent performance evaluation for defect prediction models

XUTONG LIU, SHIRAN LIU, ZHAOQIANG GUO, PENG ZHANG, YIBIAO YANG, HUIHUI LIU, HONGMIN LU, YANHUI LI, LIN CHEN, and YUMING ZHOU

State Key Laboratory for Novel Software Technology, Nanjing University

**Background.** In defect prediction community, a large number of defect prediction models have been proposed and indeed more new models are continuously being developed. **Problem.** There is no consensus on how to evaluate the performance of a newly proposed model, including the used baseline models, classification threshold setting, and performance indicators. This makes it hard, if not impossible, to measure the real progress in defect prediction by directly comparing the performance values of different models in the literature. Consequently, unintentionally misleading conclusions may be drawn on the state-of-the-art, which can cause missed opportunities for progress in defect prediction. **Objective.** We aim to propose MATTER, a fraMework towArd a consisTenT pErformance compaRison, which makes model performance directly comparable across different studies. This allows for accurately and easily keeping track of how far we are in defect prediction and discovering really useful defect prediction models, thus helping advance the state-of-the-art in a trustworthy way. **Method.** We take three actions to build a consistent evaluation framework for defect prediction models. First, we propose a simple and easy-to-use unsupervised baseline model ONE (glObal baseliNe modEl) to provide "a single point of comparison". Second, we propose using the SQA-effort-aligned threshold setting to make a fair comparison. Third, we suggest reporting the evaluation results in a unified way and provide a set of core performance indicators for this purpose, thus enabling an across-study comparison to acquire the real progress. **Result.** The experimental results show that MATTER is an effective framework to support a consistent performance evaluation for defect prediction models and hence can help determine whether a newly proposed defect prediction model is practically useful for practitioners and inform the real progress in the road of defect prediction. Furthermore, when applying MATTER to evaluate the representative defect prediction models proposed in recent years, we find that most of them (if not all) are not superior to the simple baseline model ONE in terms of the SQA-effort awareness prediction performance. This reveals that the real progress in defect prediction has been overestimated. **Conclusion.** We recommend that, in future studies, when any new defect prediction model is proposed, MATTER should be used to evaluate its actual usefulness (on the same benchmark test data sets) in order to advance scientific progress in defect prediction.

CCS CONCEPTS • Software and its engineering • Software verification and validation • Software defect analysis

**Additional Keywords and Phrases:** Defect prediction, Software quality assurance (SQA), Performance evaluation

## 1 INTRODUCTION

A defect prediction model can predict the potential locations of defects in a software project, i.e., which modules (e.g., files, classes, or functions) are defect-prone. In practice, such information is essential for software quality assurance activities. On the one hand, it enables to prioritize the modules to be inspected or tested. If high defect-prone modules are first inspected or tested, this will help find defects in a software project as early as possible. On the other hand, it enables to guide an effective allocation of inspection/testing effort. If more effort is concentrated on highly defect-prone modules, this will help find defects in a software project as many as possible under a given testing/inspection effort budget.

According to numerous publications in the last decades, our academic community has made great "progress" in defect prediction, as almost each newly proposed defect prediction model is claimed to advance the state-of-the-art [1-7]. However, several recent studies reveal that the real progress in defect prediction is not being achieved as it might have been envisaged [8][9][10].By a systematic literature review on defect prediction models published from 2000 to 2010, Hall et al. reported that, logistic regression and Naïve Bayes models, two relatively simple models, performed comparatively well compared with many other complex models [8]. Furthermore, by comparing simple module size models with complex defect prediction models published from 2002 to 2017, Zhou et al. reported that they had a prediction performance comparable or even superior to most of the existing models in the literature [9]. More recently, in 2021, Zeng et at. reported that, a simple logistic regression model outperformed deep learning models that had been demonstrated as the state-of-the-art in defect prediction, in terms of both effectiveness and efficiency on most studied projects [10]. This phenomenon reveals that, for the effectiveness in defect prediction, there is a large gap between the results reported in our academic community and the real progress made in the last decades. In other words, for most models (if not all), their effectiveness in defect prediction has been overestimated in the evaluation when they are originally proposed.

Why does our academic community tend to overestimate the effectiveness of defect prediction? One fundamental reason is that there is a lack of a consistent framework to evaluate the effectiveness of a newly proposed defect prediction model. Specifically, currently, there is no consensus among researchers on how to compare a newly proposed model against previous models, including the used baseline models, classification threshold setting, and performance indicators. First, it is common to see that different studies use different previous models (often recently published models) for comparison, as there is a lack of a simple and widely acceptable baseline model. As a result, we do not have "a global reference point" to determine the amount of real progress that is being made in defect prediction. This problem will become more serious if we consider the fact that many recently published models are often not open-source and complex (e.g., many parameters need to be tuned carefully). Therefore, they have to be re-implemented in order to enable the comparison with a newly proposed model, where a tiny implementation bias may lead to a "degraded" prediction performance. This means that a spurious advance in defect prediction may be reported in the literature, as many studies use their own re-implemented ("degraded") previous models as the baseline models to demonstrate the effectiveness of a newly proposed model. Second, within a single study, it is common to see that different models are compared using the same indicators under the SQA-effort-unaligned threshold setting. From the viewpoint of practical use, a defect prediction model is used to recommend a portion of modules in a project, which is deemed to have a high defect-proneness, for inspection or testing. In the literature, many defect prediction studies determine such modules under different efforts for different models, i.e., their comparisons are not conducted under a fair setting. Consequently, it may lead to an unfair effectiveness evaluation of a newly proposed model against the baseline models. Third, in different studies, it is common to see that different studies use different performance indicators for evaluation, as there is no agreement on what indicators should be used to measure the effectiveness in defect prediction. In particular, most performance indicators do not coincide with each other, i.e., one model may have a higher performance under indicator A but may have a lower performance under indicator B. In practice, most defect prediction studies hence select their preferred indicators (often different), if not all, to evaluate the effectiveness of a newly proposed model. Without the ongoing use of the same performance indicators, it is not possible to know the real progress in prediction performance that is being made by across-study comparison.

Clearly, for our community, the above three problems make it hard, if not impossible, to measure the real progress in defect prediction by directly comparing the performance values of different models in the literature. Consequently, unintentionally misleading conclusions may be drawn on the state-of-the-art. In this case, practitioners are likely misinformed on what is the true prediction performance of a newly proposed defect prediction model and which models are really better. Currently, it is urgent for our community to be able to find really better models for SQA engineers in practice. The fundamental of achieving this goal is to tackle how to conduct an objective evaluation on the effectiveness of a newly proposed defect prediction model. Up to now, however, the objective performance evaluation does not receive much attention in our community as it should be. To the best of our knowledge, only few studies explicitly investigated the baseline models in the field of defect prediction. In 2018, Zhou et al. [9] advocated the use of two (rather than one) baseline models, ManualDown and ManualUp, separately for non-effort-aware evaluation and effort-aware evaluation.

The problem is that, however, without the use of a single global reference point, it is hard to get a complete picture of the real progress in defect prediction. In 2019, Krishna et al. [11] proposed using the "bellwether" method to construct the baseline model against which future models should be compared. In 2022, Ni et al. proposed using "EASC" as the baseline model in defect prediction. However, due to the nature of supervised modeling, both "bellwether" and "EASC" models have a prediction performance depending on the training data used. As a result, they may produce different prediction performance values for the same target project in different studies. If they are used as the baseline models, the effectiveness of new models in different studies may be incomparable. Therefore, both "bellwether" and "EASC" models are inappropriate to be used as the global reference point to measure the real progress in defect prediction across studies. Overall, of the three important problems for an objective performance evaluation, only the baseline model has attracted (insufficient) attention, while the problems on the appropriate threshold setting and performance indicators still remain largely unexplored. For our community, in order to help advance the state-of-the-art in a trustworthy way, there is an urgent and strong need to develop a consistent performance evaluation framework for making model performance comparable across different studies. In the literature, several defect prediction frameworks have been proposed, including Song et al.'s GPF (General software defect-proneness Prediction Framework) [12] and Jing et al.'s ISDA/SSTCA+ISDA [1]. The former aims to develop a prediction model that achieves a high generalization ability by selecting a learning scheme (i.e., a combination of a data preprocessor, an attribute selector, and a learning algorithm). The latter aims to develop a prediction model that achieves a high prediction performance by dealing with the class-imbalance problem in the training data. However, they are used for model building rather than for model evaluation.

In this study, we aim to propose MATTER, a fraMework towArd a consisTenT pErformance compaRison, to achieve this goal. The core idea of our framework is to use a consistent way to calculate the benefits developers will receive in relation to their investment effort when comparing different defect prediction models. To this end, this framework consists of the following three key components: a simple yet effective baseline model (a global reference point), an SQA-effort-aligned threshold setting for cost-effectiveness evaluation (a fair performance comparison setting), and a set of core practical cost-effectiveness measures (the unified performance indicators). More specifically, we take three actions to build a consistent evaluation framework for defect prediction models. First, we propose a simple and easy-to-use unsupervised baseline model ONE to provide "a single point of comparison". Second, we propose using an SQA-effort-aligned threshold setting to conduct a fair comparison between a newly proposed defect prediction model and the baseline model. Third, we propose using a set of core practical effort-aware (i.e., considering the required SQA effort of modules as different) performance indicators to make the model evaluation in a unified way, whose ongoing use will make across-study comparison feasible. By this framework, our community can determine if a newly proposed defect prediction model is practically useful for practitioners and how far we really are in the road of defect prediction.

In this study, we make the following contributions:

- We conduct a systematic analysis of the pitfalls of the existing approaches for evaluating the effectiveness of a newly proposed defect prediction model. We show that, for a new model, the existing evaluation approach may lead to a biased evaluation due to the lack of the ongoing use of a global baseline model, a fair performance computation setting, and unified performance indicators. As a result, the reported progress in defect prediction in the literature may be not reliable or even spurious. This may lead to unintentionally misleading conclusions on advancing the state-of-the-art, thus possibly causing missed opportunities for progress in defect prediction.
- We propose MATTER, a fraMework towArd a consisTenT pErformance compaRison, for the evaluation of a newly proposed defect prediction model. Different from the existing evaluation approach, our framework proposes the ongoing use of a simple yet effective baseline model, an SQA-effort-aligned threshold setting, and unified practical cost-effectiveness performance indicators. By MATTER, on the one hand, researchers can determine whether a newly proposed defect prediction model is practically essential for practitioners. On the other hand, more importantly, we can acquire real progress in defect prediction, which can promote our defect prediction community a steady and healthy development. To the best of our knowledge, this is the first time to systematically take into account such a problem and propose a solution in the defect prediction community.

- We use MATTER to investigate the real progress in defect prediction. Surprisingly, we find that, when the SQA-effort required is considered, our simple baseline model has a prediction performance comparable or even superior to most of the investigated representative defect prediction models reported in the literature. On the one hand, the results caution us that the real progress in defect prediction is not being achieved as it might have been envisaged. On the other hand, more importantly, the results demonstrate the usefulness of our framework in acquiring the real progress in defect prediction.
- We provide the source code for this study, which can be easily used in future defect prediction studies. This will facilitate researchers to use our framework MATTER to evaluate the effectiveness of a newly proposed model, thus helping develop really effective defect prediction models.

The remainder of this paper is organized as follows. Section 2 introduces the relevant background, including defect prediction models and defect prediction scenarios. Section 3 analyzes the pitfalls in the existing model evaluation. Section 4 presents the framework MATTER for consistent model evaluation. Section 5 describes the experimental setup. Section 6 reports the experimental results in detail. Section 7 discusses the experimental results. Section 8 presents the implications. Section 9 summarizes the related work. Section 10 analyzes the threats to the validity of our study. Section 11 concludes the paper and outlines directions for future work.

## 2 BACKGROUND

In this section, we introduce the background of software defect prediction, including defect prediction scenarios and defect prediction models.

### 2.1 Defect prediction scenarios

In real-world software development, software quality assurance (SQA) resources are often limited [13]. As a result, not all the modules in a target project can be inspected/tested. For a given software project, SQA engineers hence have to select a portion of modules from the project for inspection or testing. Consequently, the following problem is naturally raised: which modules should be selected for quality assurance? According to the literature, defect prediction is a promising module selection method: first, predict the defect-proneness of each module in the project; then, select those modules with a large defect-proneness for quality assurance. In this context, classification and ranking are the most studied defect prediction scenarios.

In the classification scenario, a threshold is first set to classify the modules in a target project into two categories: "defective" and "clean". If a module has a predicted defect-proneness higher than the threshold, it will be classified as "defective"; otherwise, it will be classified as "clean". In nature, the module selection problem is treated as a binary classification problem. After that, only those modules classified as "defective" will be recommended to SQA engineers for quality assurance. In this scenario, one implicit assumption is that SQA resources are enough to inspect/test all the modules that are classified as "defective", i.e., all the predicted "defective" modules will be inspected/tested. In particular, no particular module order is assumed, when these modules are inspected/tested. The expectation is that an effective classification can help SQA engineers find as many defects as possible under a reasonable precision.

In the ranking scenario, the modules in a target project are ranked from the highest to the lowest predicted defect-proneness. In nature, the module selection problem is treated as a ranking problem. Given such a ranking list, SQA engineers can select the modules from the top to the low rank for quality assurance until the available inspection/testing resources are exhausted. Unlike the classification scenario, the ranking scenario does not make any particular assumption on the amount of available resources. In practice, under a given budget of testing/inspection effort, it is expected that an effective module rank can help SQA engineers find the defects in the target project as early as possible and as many as possible.

## 2.2 Defect prediction models

In order to build a defect prediction model, the current practice is to leverage software metrics that quantify the complexity of code [14-19], development process [19-25], or even development organization as the predictors [26,27]. The rationale behind this is that, the more complex these factors are, the more likely a defect will be introduced into the corresponding modules. In the defect prediction community, supervised and unsupervised techniques are the two most commonly used modeling techniques for defect prediction models.

For supervised defect prediction models, the goal is to learn a function (either explicit or implicit) from the labeled training data that best approximates the relationship between software metrics and defect-proneness observed in the data. The advantage of supervised defect models is that previous knowledge between predictors and defects can be learned and hence it is possible to produce an accurate prediction on unseen data. In the past decades, a large number of supervised defect prediction models have been proposed, ranging from traditional regression-based [28], rule-based [29], and tree-based [30,31] to modern deep-learning-based models [32]. Overall, they have shown a promising defect prediction performance. The main disadvantage of supervised defect models, however, is that labeled training data may be expensive to collect or even unavailable.

Unlike supervised models, unsupervised defect prediction models do not need labeled training data. The goal of unsupervised models is to find the structure and patterns from unlabeled data. In the literature, unsupervised defect prediction models mainly include association-based [9] and clustering-based [33,34]. Compared with supervised models, unsupervised defect prediction models often have a lower model building cost, as there is no need to use the label information. Since unsupervised learning does not use any supervision, it is expected that they may give less accurate prediction results compared with supervised learning.

## 3 PITFALLS IN EXISTING MODEL EVALUATION

When a new prediction model is proposed, it is necessary to evaluate its effectiveness in defect prediction, i.e., the ability to help developers find defects in a project. Fig. 1 describes the general model evaluation procedure used in the literature. First, determine a number of previous defect prediction models with which the newly proposed defect prediction model will be compared. In the current practice, it is common to choose classic defect prediction models and/or state-of-the-art (i.e., recently published) defect prediction models as the baselines. Second, for each pair of training and test sets, apply those models built on the same training set to the same test set for calculating the prediction performance values. For each model, it is common to use a default value as the classification threshold. The most commonly used performance indicators include recall, precision, F1 (the harmonic mean of the precision and recall), G1 (the harmonic mean of the recall and 1-PF), AUC (area under ROC), and MCC (Matthews Correlation Coefficient) [9,35]. Third, after obtaining their prediction performance values, employ a statistical method (e.g., the Wilcoxon signed-rank test [36]) to examine whether the newly proposed model is significantly better than the baselines. Furthermore, compute the effect size (e.g., Cliff's delta [37]) to determine whether their difference is practically important. To date, hundreds of defect prediction models have been evaluated in this way in the literature [38-42].

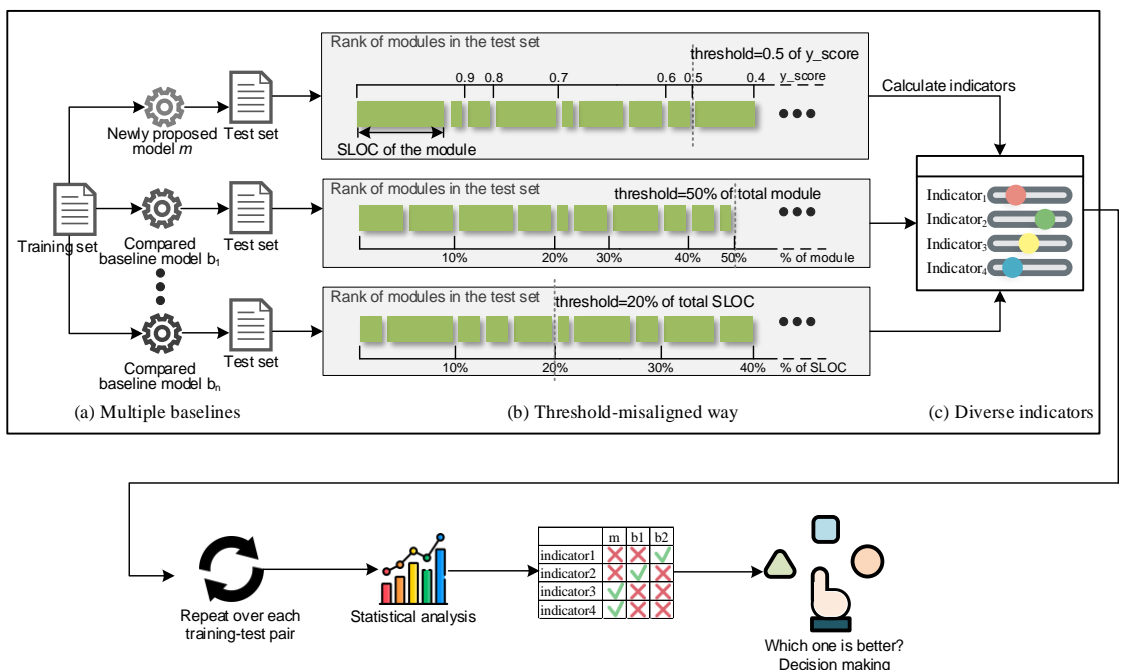

Figure 1: The general evaluation procedure of a newly proposed defect prediction model

However, as mentioned in section 1, we find that there are three problems when taking a close look at the application of the above model evaluation process in the literature: (1) the lack of a common "stable" baseline model as the global reference point, (2) the unfair performance comparison across models due to the use of SQA-effort unaligned thresholds to identify potentially defective modules, and (3) the unfeasible performance comparison across studies due to the lack of the ongoing use of the same performance indicators. Due to the above three problems, our community indeed is unable to know how far we have advanced in the journey, although a large number of defect prediction studies have been published. In particular, currently, there is an increasing interest in defect prediction, leading to more and more models being published. If we do not use a consistent framework to conduct model performance evaluation, it is very likely to see false progress in defect prediction. In this case, we may not only waste our precious research resources but also miss opportunities for advancing defect prediction, which is harmful to our community.

### 3.1 Lack of a common "stable" baseline model as the global reference point

In order to get an understanding of how serious this problem is, we search for representative defect prediction studies with the following filtering conditions: (1) they must be published in the premier journals ACM Transactions on Software Engineering and Methodology (TOSEM) or IEEE Transactions on Software Engineering (TSE); (2) they should be published within the recent six years (2017~2022); and (3) they aimed to propose new defect prediction models. Condition 1 ensures that the included studies have a potentially high impact [43,44]. Condition 2 ensures that the included studies represent the state-of-the-art, while condition 3 ensures that the focus is a new defect prediction model. With the above search criteria, consequently, we find 11 defect prediction studies.

Table 1. The baseline models used in recent studies that are published in TOSEM and TSE within the latest six years (2017~2022) and aim to propose a new defect prediction model

| | SSTCA+ISDA [1] | HDP-KS [45] | Bellwether [11] | MSMDA [2] | DBN [3] | FENCES [4] | DESCo [7] | top-core [47] | KSETE [5] | EASC [46] | TDTSR [6] |
|---|---|---|---|---|---|---|---|---|---|---|---|
| | 2017 | 2018 | 2019 | 2019 | 2020 | 2020 | 2020 | 2021 | 2021 | 2022 | 2022 |
| CCA+ | | | | x | | | | | x | | |
| HDP-KS | | | | x | | | | | x | | x |
| CTKCCA | | | | | | | | | x | | x |
| TSEL | | | | | | | | | | | x |
| NNFilter | | | | x | | | | | x | | |
| TNB | | | x | | | | | | x | | |
| TCA+ | x | | x | x | x | | | | x | | |
| HISNN | | | | | | | | | x | | |
| BellWether | | | | | | | | | x | | |
| Ree | | | | | | | | x | | | |
| Ree+BW | | | | | | | | x | | | |
| Ree+PR | | | | | | | | x | | | |
| Ree+degree | | | | | | | | x | | | |
| CamargoCruz09-DT | | | | | | | | | | x | |
| Menzies11-RF | | | | | | | | | | x | |
| Turhan09-DT | | | | | | | | | | x | |
| Watanabe08-DT | | | | | | | | | | x | |
| ManualDown | | | | | | | | | | x | |
| ManualUp | | | | | | | | | | x | |
| RH | | | | | | x | | | | | |
| MO | | | | | | x | | | | | |
| SCC | | | | | | x | | | | | |
| CK | | | | | x | x | | | | | |
| AST | | | | | x | | | | | | |
| DBN | | | | | | | x | | | | |
| CPDP-IFS | | x | | x | | | | | | | |
| MS-TrAdaBoost | | | | x | | | | | | | |
| HYDRA | | | | x | | | | | | | |
| SC | | | | x | | | | | | | |
| CLIFF+MORPH | | | | x | | | | | | | |
| VCB | x | | x | | | | | | | | |
| CLAMI | | x | | | | | | | | | |
| CPDP-CM | | x | | | | | | | | | |

Table 1 summarizes for these 11 defect prediction studies [1-7,11,45-47] the used baseline models. In this table, each column reports a newly proposed defect prediction model, including the name, the year it was published, and the corresponding study it was proposed. The rows report the baseline models used when evaluating the effectiveness of these newly proposed defect prediction models. A cell with "x" indicates that, in the corresponding study, the newly proposed model was compared against the baseline model in the corresponding row. In particular, "x" marked in blue indicates that the corresponding baseline model is the most frequently used baseline model, while "x" marked in black indicates that the corresponding baseline model is used in only one study. In the other cases, "x" is marked in red.

From Table 1, we can see that more than 30 baseline models are involved in the 11 defect prediction studies. However, there is no single common baseline model used in all the 11 studies. Indeed, in each study, the authors used their preferred prior models as the baseline models to investigate the effectiveness of the newly proposed model. This is especially true for 2020-EASC [46] and 2021-top-core [47], as the used baseline models do not overlap with those used in the other studies. In each of the 11 studies, the authors stated that their newly proposed model outperformed the baseline models and hence advanced the state-of-the-art. But, due to the lack of a common baseline model, we are unable to determine the relative effectiveness of those 11 new models. More importantly, due to the lack of a global reference point, we are unable to know how far we really have come in the journey of defect prediction.

The above problem is even worse if we take into account the fact that many prior supervised models are complex (e.g., many parameters need to be tuned carefully) and not open-source. As a result, subsequent studies have to re-implement them in order to use them as the baseline models. However, a tiny implementation bias may lead to a "degraded" prediction performance. Therefore, it is possible to report a spurious advance when a newly proposed model is shown to have a higher effectiveness than these baseline models. Furthermore, different subsequent studies may have different implementations and use different training data for the same prior models. Consequently, *on the same test data, the same baseline models in different studies may exhibit considerably different performance values*. Table 2 summarizes the mean G1 values on the same five target projects in data set AEEEM (EQ, JDT, Lucene, Mylyn, and PDE) for the baseline models in different studies. In this table, the second column reports for each new model the mean G1 value when the model is proposed in the original study. The third to ninth columns report the mean G1 values for the chosen baseline models. From Table 2, we can see that all the baseline models have very unstable performance values across studies. For example, TCA+ [48] has a G1 of only 0.283 in 2021-KSETE [5] but has a considerably larger G1 in the other three studies. In other words, in different studies, TCA+ exhibits very different performance values on the same test data set. In this case, TCA+ indeed loses the role that a baseline model purports to play, as we cannot use TCA+ as a global reference point to make a meaningful comparison of G1 across studies for KSETE [5], MSMDA [2], Bellwether [11], and ISDA/SSTCA+ISDA [1]. The key message from the above fact is that we cannot use a supervised model as a global reference point to determine the real progress in defect prediction even if it is used as the common baseline model in all studies. A more appropriate baseline model that can provide a stable "global reference" is needed to be proposed.

Table 2. The mean G1 values on the same five target projects in AEEEM for the newly proposed defect prediction models as well as the used baseline models in different studies

| New models | G1 | Baseline models | | | | | | |
| --- | --- | --- | --- | --- | --- | --- | --- | --- |
| | | TCA+ | CCA+ | HDP-KS | B.Filter | TNB | VCB | Bellwether |
| KSETE(2021) | 0.653 | 0.283 | 0.278 | 0.024 | 0.339 | 0.406 | | 0.478 |
| MSMDA(2019) | 0.696 | 0.621 | 0.615 | 0.613 | 0.593 | | | |
| Bellwether(2019) | 0.750 | 0.688 | | | | 0.654 | 0.558 | 0.750 |
| SSTCA+ISDA(2017) | 0.771 | 0.521 | | | | | 0.638 | |

## 3.2 Unfair across-model comparison due to SQA-effort-unaligned threshold setting

In practice, SQA effort is often limited when inspecting/testing the potentially defective modules identified by defect prediction models. Therefore, SQA engineers hope to detect as many real defects as possible with the limited SQA effort. According to the literature [49], the SQA effort involved in defect prediction mainly consists of code inspection/testing effort and context-switch effort. Given a set of potentially defective modules, SQA engineers need to inspect/test these modules to examine whether they contain real defects. To this end, SQA engineers need to check the code of each module and switch from one module to another module. During this procedure, on the one hand, the larger a module is, the more code inspection/testing effort it tends to consume. On the other hand, the more modules are examined, the more context-switch effort it needs to spend. Considering the above facts, it is important to align the SQA effort when comparing the effectiveness of different defect prediction models. Otherwise, misleading conclusions may be drawn.

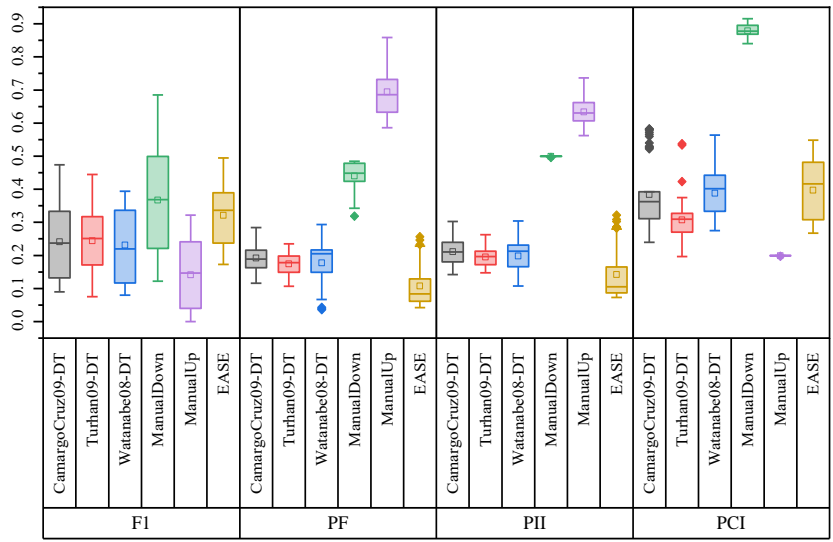

Figure 2: The comparison of EASC with the baseline models on the MA-SZZ-2020 data set under the SQA-effort-unaligned threshold setting

In the current practice, however, it is common to use default threshold settings, which are often SQA-effort-unaligned (i.e., the SQA-effort for inspecting/testing the predicted defect-prone modules is not the same for different prediction models), to conduct model performance comparisons. For example, as shown in Table 1, when EASC is proposed [46], it is compared with six baseline models: four supervised models (CamargoCruz09-DT [50], Menzies11-RF [51], Turhan09-DT [52], and Watanabe08-DT [53]) and two unsupervised models (ManualDown and ManualUp [9]). During the comparison, all the supervised models (including EASC) use the default threshold of 0.5 to classify the modules in a project: a module is classified as "defective" if it has a probability of being defective larger than 0.5; otherwise, it is classified as "clean". However, ManualDown uses 50% as the classification threshold (the top 50% largest modules are classified as "defective"), while ManualUp uses 20% as the classification threshold (the top smallest modules accounting for 20% of total SLOC are classified as "defective"). Fig. 2 reports the comparison of EASC with the baseline models on a quality defect dataset named MA-SZZ-2020 [54], which is generated by the state-of-the-art SZZ variant MA-SZZ [55], under such an SQA-effort-unaligned threshold setting. Here, PF denotes the probability of false alarm, PII denotes the Percent of Instances (i.e., one instance is one module in our context) Inspected/tested [46], and PCI denotes the Percent of Code Inspected/tested. According to F1 and PF, EASC exhibits a competitive prediction performance. In particular, when compared with ManualDown, EASC exhibits the following prominent advantages: a comparable F1, a considerably lower PF, a considerably lower PII, and a considerably lower PCI. It seems that EASC enables SQA engineers to spend much lower SQA effort to achieve a competitive effectiveness. From the above experimental results, it is accordingly concluded that ESAC is superior to ManualDown.

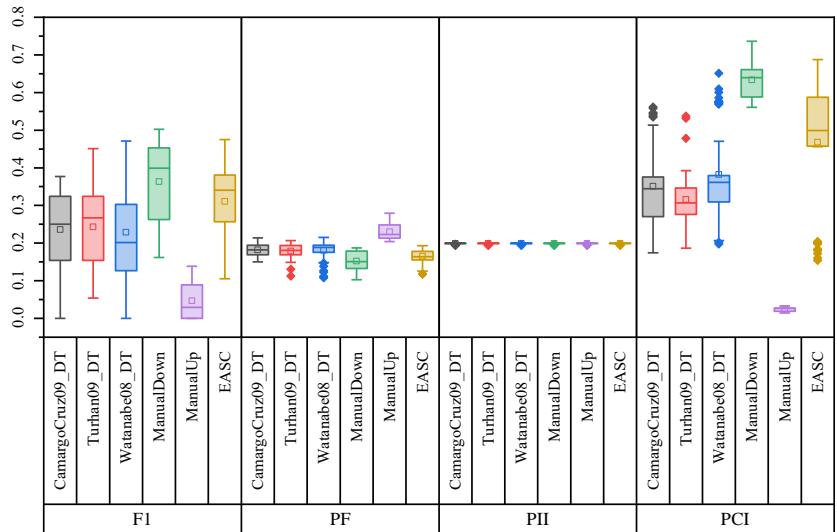

Figure 3: The comparison of EASC with the baseline models on the MA-SZZ-2020 data set under the PII-aligned threshold setting (PII = 20%)

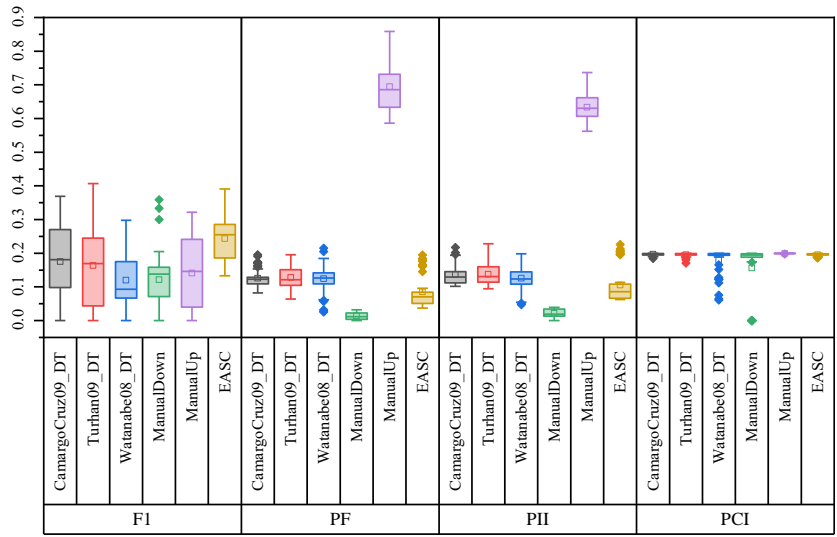

Figure 4: The comparison of EASC with the baseline models on the MA-SZZ-2020 data set under the PCI-aligned threshold settings (PCI = 20%)

However, the above conclusion is based on the SQA-effort-unaligned threshold settings, making the across-model performance comparison unfair. In practice, for two defect prediction models, SQA engineers would like to know which one is superior under the same SQA effort. In other words, given the same SQA effort, which model can help them to find more defects? Fig. 3 reports the comparison of EASC with the baseline models on the MA-SZZ-2020 data set when the percent of modules examined is fixed to 20%. When comparing ManualDown against EASC, we can see that: ManualDown exhibits a slightly higher F1, a slightly lower PF, and a slightly higher PCI. This means that ManualDown is competitive with EASC. Fig. 4 reports the comparison of EASC with the baseline models on the MA-SZZ-2020 data set when the percent of code examined is fixed to 20%. When comparing ManualDown against EASC, we can see that: although ManualDown has a lower F1, PF and PII are also lower. This means that ManualDown is still competitive with EASC when we consider the cost-effectiveness (i.e., the number of defects found under each unit effort).

Overall, the above results reveal that the SQA-effort-unaligned threshold setting can lead to unintentionally misleading conclusions for model performance evaluation. In fact, we find that this problem exists in the nine studies shown in Table 1. The two exceptions are 2018-HDP-KS [45] and 2021-top-core [47], which respectively use AUC and $P_{opt}$ (threshold-independent indicators) to conduct the performance evaluation. However, AUC and $P_{opt}$ are two global performance indicators [56,57] computed on the whole rank on the test set. The threshold setting problem cannot be avoided by using those global performance indicators if we consider the fact that the purpose of a defect prediction model is to recommend a portion of modules from a software project for examination (if all the modules in a project are recommended, we do not need defect prediction). In this sense, the SQA-effort-unaligned threshold setting problem is common in defect prediction studies, which demands immediate attention in our community. Otherwise, the conclusions from performance comparison may mislead both researchers and practitioners.

### 3.3 Unfeasible across-study comparison due to diverse performance indicators

In order to determine the real progress in defect prediction, we need to be able to make a comparison of model prediction performance across different studies. To this end, a basic requirement for across-study comparison is using the same test sets. However, even if different studies use the same tests sets, to make across-study comparison feasible, except the need for a common baseline model and the SQA-effort-aligned threshold setting, there is also a need to use the same prediction performance indicators across different studies. In the literature, within a single study, it is natural to use the same indicators to compare the performance of different defect prediction models. However, in different studies, different authors tend to use their preferred performance indicators. As a result, diverse performance indicators, often different or even conflicting, are used in different studies. In this context, it is unfeasible to make an across-study performance comparison for the proposed defect prediction models in different studies.

Table 3 summarizes the performance indicators used in the 11 representative defect prediction studies shown in Table 1. In this table, a cell with "x" means that a study (indicated by the column) uses the corresponding indicator (indicated by the row) for model performance evaluation. As can be seen, more than 15 indicators are involved in these 11 defect prediction studies. In particular, there is no single common performance indicator used in all the 11 studies. Consequently, we are unable to make a direct performance comparison for the proposed eleven new defect prediction models even if they use the same test sets. Furthermore, it is often not possible to infer a common indicator from the diverse indicators in different studies for an across-study model performance comparison. The reason is that these studies do not provide sufficient information for this purpose. For example, if we want to use F1 as the common indicator to conduct a performance comparison for these new models, there is a need to infer the F1 indicator for those studies that do not report F1. As shown in Table 3, of the 11 studies, six studies do not report F1. For these six studies, we cannot infer F1 from the reported indicators due to insufficient information (i.e., recall and pf on test sets). The above problem makes it impossible to compare the F1 values across the eleven new models.

In the literature, a large number of performance indicators have been proposed to measure model performance. These indicators measure the performance of a model from different viewpoints, often resulting in a conflicting conclusion on which model is better. For example, a model is shown to be better than another model in terms of indicator A, while the conclusion may be opposite in terms of another indicator B. This is one possible reason why prior studies use their preferred indicators to conduct performance evaluations. However, the above practice hinders across-study model comparison. As a result, when a new model is shown to be superior to prior models in a study, we are unable to know whether it really advances the state-of-the-art due to the inability of the model comparison across studies. To tackle the above problem, there is a need to use the same appropriate performance indicators across different studies. Such indicators should be appropriate to evaluate the practical effectiveness of models in a real-world scenario. However, this is challenging, as different indicators serve different purposes and there is no consensus on which indicators should be used.

Table 3. The performance indicators used in recent studies that are published in TOSEM and TSE within the latest six years (2017~2022) and aim to propose a new defect prediction model

| Indicator | SSTCA+ISDA [1] | HDP-KS [45] | Bellwether [11] | MSMDA [2] | DBN [3] | FENCES [4] | DESCo [7] | top-core [47] | KSETE [5] | EASC [46] | TDTSR [6] |
|---|---|---|---|---|---|---|---|---|---|---|---|
|  | 2017 | 2018 | 2019 | 2019 | 2020 | 2020 | 2020 | 2021 | 2021 | 2022 | 2022 |
| precision |  |  |  |  | x | x |  |  |  |  | x |
| recall | x |  |  |  | x | x |  |  | x | x | x |
| PF | x |  |  |  |  |  |  |  | x |  | x |
| F1 |  |  |  |  | x | x | x |  |  | x | x |
| F4 | x |  |  |  |  |  |  |  |  |  |  |
| G1 |  |  | x | x |  |  |  |  | x |  |  |
| MCC |  |  |  |  |  |  |  |  | x |  |  |
| AUC | x | x |  | x |  | x | x |  | x | x | x |
| recall@20%LOC |  |  |  |  | x |  |  |  | x | x |  |
| IFA |  |  |  |  |  |  |  |  | x | x |  |
| Popt |  |  |  |  |  |  |  | x |  | x |  |
| PII@20%LOC |  |  |  |  |  |  |  |  |  | x |  |
| PII@1000LOC |  |  |  |  |  |  |  |  |  | x |  |
| PII@2000LOC |  |  |  |  |  |  |  |  |  | x |  |
| recall@1000LOC |  |  |  |  |  |  |  |  |  | x |  |
| recall@2000LOC |  |  |  |  |  |  |  |  |  | x |  |

## 4 MATTER: A FRAMEWORK TOWARD A CONSISTENT PERFORMANCE COMPARISON

In this section, we propose MATTER, a fraMework towArd a consisTenT pErformance compaRison for defect prediction models. At the high level, MATTER consists of three components: one coMmon bAseline, two efforT-aligned Threshold settings, and three corE indicatoRs. Specifically, we propose to use ONE (glObal baseliNe modEl), an easy-to-use baseline model, as "the global reference point" to evaluate whether a newly proposed model *model* is effective in defect prediction. Furthermore, we propose to use two SQA-effort-aligned settings to make a fair performance comparison between *model* and ONE. In particular, we advocate that all studies include the same three core performance indicators to conduct the performance evaluation and report the detail prediction result on each test set, thus enabling a feasible across-study comparison. By continuously applying MATTER to the same benchmark test data sets in different studies, our community can achieve a consistent performance evaluation for defect prediction models. As a result, researchers are able to find practically effective prediction models for SQA engineers and hence can help advance the state-of-the-art in software defect prediction in a trustworthy way.

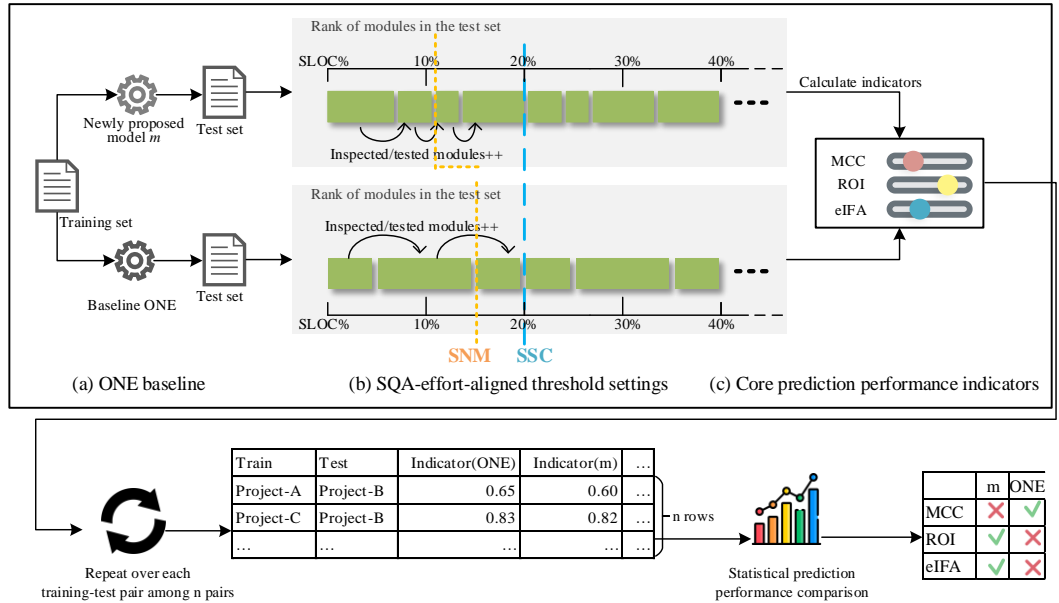

Figure 5: MATTER: A framework for consistent comparisons between defect prediction models

Fig. 5 shows the evaluation workflow in our framework MATTER. As can be seen, when a new defect prediction model $m$ is proposed, the following steps are employed to evaluate its prediction performance, informing whether it really advances the state-of-the-art in defect prediction.

(1) Generate defect-proneness rankings for $m$ and ONE. On a given test set, for each model, rank the modules in the corresponding target project from the most to the least predicted defect-proneness. As a result, two rankings are obtained, one for model $m$ and another for ONE;

(2) Apply two SQA-effort-aligned threshold settings. For $m$ and ONE, use the following two threshold settings to obtain the corresponding classification results: SNM (Same Number of Modules) and SSC (Same Size of Code). The purpose of this is to enable a comparison under the same inspected/tested number of modules (SNM) or the same inspected/tested code size (SSC).

(3) Calculate three core prediction performance indicators. For each model, after obtaining the classification result, compute the following performance indicators: MCC, ROI (Return Over Investment, i.e., the number of defective modules found per unit SQA-effort), and eIFA (SQA-effort required in Initial False Alarm).

(4) Repeat (1)~(3) over each training-test pair to obtain the performance values. Consequently, if there are $n$ training-test pairs, for each indicator, both $m$ and ONE will have $n$ prediction performance values.

(5) Conduct a statistical prediction performance comparison. Under each indicator, use a statistical method (say the non-parametric Scott-Knott ESD test [58]) to examine if $m$ is significantly better than ONE. If there is a significant difference at the significance level of 0.05, examine whether the effect size (say Cliff's delta [37]) is non-trivial, i.e., whether their difference is of practical importance.

In the following, we will introduce how to design ONE to make it appropriate for being a global reference point (Section 4.1), what is the rationale behind the SNM and SSC effort-alignment threshold settings (Section 4.2), and why MCC, ROI, and eIFA matter for consistent model performance comparison (Section 4.3).

### 4.1 Using a simple, strong, and stable baseline model as "the global reference point"

Our purpose is to provide a framework MATTER for consistent model performance comparison, thus informing whether a newly proposed defect prediction model really advances the state-of-the-art. To this end, the baseline model in MATTER should have the following good properties:

- Simple in implementation. The baseline model should be simple enough to make researchers easy to implement without any bias [59]. Ideally, a reference implementation should be publicly available.
- Strong in prediction ability. The baseline model should be strong enough in prediction to provide a "floor performance", helping quickly filter out any model that falls "below the floor" [11].
- Stable in prediction performance. The baseline model should have a stable prediction performance across studies, making it possible to be "a global reference point" [11,31,59].

Without the 3S properties (Simple, Strong, and Stable), a model is inappropriate to be used as the common global baseline model for consistent model performance evaluation.

In MATTER, ONE (glObal baseliNe modEl) is designed to adhere to the 3S properties. In particular, we hope that, with the help of ONE, SQA engineers could find the defects in a target project as quickly as possible and as many as possible per unit SQA effort. The use of such a ONE as the baseline model will help select out practically useful defect prediction models. Therefore, we design ONE based on the following intuitions on the relationship between size and defect-proneness at the module level: (1) larger modules are more likely to contain defects; (2) very large modules may have a very low defect density; and (3) smaller modules are more likely to have higher defect densities. Based on the above intuitions, ONE uses the following steps to generate a defect-proneness ranking for the $N$ modules in a target project.

- **Step 1: rank the $N$ modules in descending order according to their module size.** This step is based on the first intuition. In order to make it easy to apply, ONE uses the module size measured by SLOC (Source Lines Of Code) to rank the modules. The reasons are two-folds. First, many prior studies show that module size measured by SLOC has a high positive correlation with defect-proneness [9,60-66]. Second, SLOC is the most commonly used module size metric, which is cheap to collect.
- **Step 2: put the top largest module set E at the bottom and rank them in ascending order.** This step is based on the second and third intuitions. Let $E$ be the set of the top largest modules. First, put $E$ at the bottom of the ranking, as they tend to have a very low defect density (the second intuition). Second, sort the modules within $E$ from the smallest to largest module size, hoping that the modules with a higher defect density will be first inspected/tested (the third intuition). In ONE, a parameter "excluded-code-size-percentage" is used to determine $E$: the modules in $E$ account for the "excluded-code-size-percentage" percent of the total project size. The default value of "excluded-code-size-percentage" is set to 20%.
- **Step 3: classify the modules in N by the classification threshold.** Given a classification threshold $\tau$, the modules in $N$ are classified into two parts: the top predicted buggy part and the bottom predicted clean part. In default, $\tau$ is set so that the modules in $D$ account for 20% of the total module number or the total code size. The reason for this default setting is that it has been reported that the distribution of faults in a software project in general exhibits a Pareto distribution, i.e., 80% faults concentrate on 20% modules (code size) of the project [67-69]. By setting 20% as the default value, we expect that most faults in the target project will be discovered by inspecting/testing the predicted buggy modules.

Fig. 6 shows the process in ONE to generate a module ranking for a target project, which corresponding to step 1 and step 2. As can be seen, the resulting ranking consists of two parts: $R$ (i.e., $N$-$E$, those modules with moderate to small size), and E (very large modules). At the high level, ONE prioritizes those modules with moderate size, rather than small modules or very large modules, in a project for inspection/testing. As a result, ONE does not fall into the following common traps: (1) the high context switching effort needed by inspecting/testing many small modules; and (2) the high code inspection/testing effort needed by very large modules. Therefore, intuitively, ONE can lead to a cost-effective defect prediction, i.e., if we follow the module ranking given by ONE to inspect/test, the defects would be found reasonably quickly, and a reasonable number of defects would be found per unit SQA effort. In this sense, ONE has the potential to be a good baseline for defect prediction models.

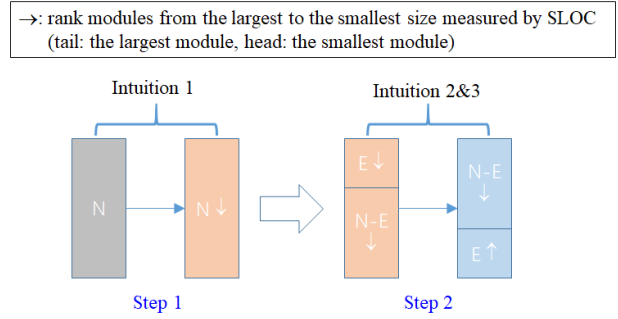

Figure 6: The process in ONE to generate a module ranking for a target project

---

ALGORITHM 1: ONE (glObal baseliNe modEl), the baseline model in MATTER

**Input:**
*N*: a list of modules in the test project
*excluded-code-size-percentage*: the percentage of the total code size for determining *E* (the default value = 20%)
*classification threshold $\tau$*: the threshold used to classify modules as defective or clean (the default value = 20%)

**Output:**
*PM*: the value of performance metrics;
*default_classification_result*: default classification result contains binary prediction label of each module ("defective" as 1 and "clean" as 0)

1: *N* = **rank**(*N*, descending=True);
// split N into two sublists: *E* (the top largest modules accounting for the "excluded-code-size-percentage"
// percent of the total code size) and *R* (the remaining modules)
2: (*E*, *R*) = **split**(*N*, length=*excluded-code-size-percentage*);
3: *E* = **rank**(*E*, descending=False);
// get the final rank by appending *E* to the end of *R*
4: *N* = **catenate**(*R*, *E*);
5: *PM* = **calculatePerformance**(*N*, $\tau$);
6: **return** *PM*.

---

The pseudo-code of ONE is presented in Algorithm 1. First, rank the *N* modules in descending order according to their code size (Line 1). Second, split *N* into two parts (Line 2): *E* (the top largest modules accounting for the "excluded-code-size-percentage" percent of the total code size) and *R* (the remaining modules). Third, rank the modules in *E* in ascending order according to their module size and obtain *N* by appending *E* to the end of *R* (Lines 3-4). Fourth, calculate the values of performance metrics with *N* under the classification threshold $\tau$ (Line 5). Fifth, return the result *PM* (Line 6). As can be seen, ONE only leverages the module size as measured by SLOC in a test set to predict defect-proneness. In nature, ONE is an unsupervised model based on the test set, without the need for any training data. Due to the low data collection cost and the low model building cost, ONE is *simple enough* for researchers to implement while not introducing implementation bias. Furthermore, as will be investigated in Section 6, ONE is competitive or even superior to most representative defect prediction models and hence is *strong enough* to be used for selecting out practically useful prediction models. In particular, ONE has the *same* (i.e., *stable*) prediction performance on the same test set in different studies, independent from the training sets. In this sense, ONE can provide a global reference point, making it possible to conduct a cross-study model performance comparison. Therefore, the use of ONE as a baseline model satisfies all the 3S properties.

### 4.2 Using the SQA-effort-aligned threshold setting to make a fair performance comparison

How to align the required SQA effort for a fair cross-model prediction performance comparison? As stated in Section 3.2, for each defect prediction model, the total SQA effort mainly consists of two types of efforts: context switching effort (indicated by the number of inspected/tested modules) and code inspection/testing effort (indicated by the inspected code size). Ideally, we should align the total SQA effort for compared models. However, in practice, it is hard to make such a strict alignment, because it is often not possible that two models recommend the exact same number of modules with the exact same total code size. One compromise solution to this problem is to relax the total SQA effort alignment to a partial SQA effort alignment: align only the code inspection/testing effort or align only the context switching effort. Indeed, given two models, in real-world development, it is natural to ask the following two questions: what if an SQA engineer inspects/tests the same number of modules according to their recommendations? what if an SQA engineer inspects/tests the modules with the same code size according to their recommendations? As can be seen, the partial SQA effort alignment is consistent with the current defect prediction practice. In nature, the former aims to conduct a cross-model comparison under the alignment of context switching effort, while the latter aims to conduct a cross-model comparison under the alignment of code inspection/testing effort.

When applying MATTER to evaluate the effectiveness of a new model *model*, we hence recommend using the partial SQA-effort-aligned threshold setting to make a fair comparison between *model* and ONE. More specifically, we advocate using one of the following two ways to adjust their classification thresholds for the partial SQA effort alignment.

- SNM (the Same Number of inspected/tested Modules). Adjust the classification threshold of ONE so that it recommends the same number of "defective" modules as that recommended by *model*. The purpose of this is to align their context switching efforts.
- SSC (the Same Size of inspected/tested Code). Adjust the classification threshold of ONE so that it recommends the "defective" modules with the same code size as that recommended by *model*. The purpose of this is to align their code inspection/testing efforts.

By the comparison with ONE under the SNM/SSC threshold settings, we are able to determine whether the model *model* is superior to the simple baseline. However, one inherent problem is that the threshold used in ONE is indeed determined based on the threshold used in the new model. If many new models are evaluated in such a way, each of them will be compared against ONE with different thresholds. As a result, we are unable to use ONE as the global reference point to determine which new model is more effective. In order to tackle this problem, we suggest that all the models, including ONE, are compared under the same SNM and the same SSC: 20% Percent of Instances Inspected/tested (PII) for SNM and 20% Percent of Code Inspected/tested for SSC. The rationale behind 20% is the Pareto principle [70,71]: 20% of the modules or code size account for 80% of defects in a project. Indeed, it has been a practice to inspect 20% of the modules or code size in practice, hoping that most defects would be found [9,67,68,70,72-74]. By consistently using SNM (PII=20%) and SSC (PCI=20%) across different studies, we not only can make fair across-model comparisons within one single study, but also can make it possible to fairly compare different models across studies. This enables us to use ONE as a global reference point to determine the real advance in defect prediction. In Section 7.1.1, we will analyze the impact of the values of PII and PCI on the model performance comparison.

### 4.3 Using three core indicators as a starting point for consistent performance comparison

Given a newly proposed model *model*, there is a need to evaluate how well it performs compared with the baseline model "ONE" under the SQA-effort-aligned threshold setting. In this context, which performance indicators should be used for comparison? Since SQA practitioners are the users of defect prediction models, it is important to evaluate and select models from the viewpoint of real-world software development. For a defect prediction model, practitioners are particularly interested in knowing the following two aspects. On the one hand, whether *model* will lead to a higher ROI (**R**eturn **O**ver **I**nvest, i.e., the number of defective modules found per unit SQA-effort) than ONE? If *model* has a higher ROI, this means that more defects would be found when spending the same SQA-effort compared with ONE. On the other hand, whether *model* will waste less unnecessary effort on non-defective instances before the first actual defective module is found? In real-world software quality assurance, practitioners may stop inspecting/testing when they could not get promising results (i.e., find defective instances) within the first few inspected instances. If *model* has a lower SQA-effort involved in eIFA (SQA-**e**ffort required in **I**nitial **F**alse **A**larms), this means that the first actual defective module will be found more quickly compared with ONE. In view of the above facts, we recommend ROI and eIFA to be two core indicators for consistent model performance comparison in defect prediction.

One prominent advantage of ROI and eIFA is that they can be calculated at the model application stage (i.e., the real-world defect prediction scenario where the ground-truth labels of the modules in a target project are unavailable in advance) as well as at the model evaluation stage (i.e., the model is evaluated on test sets where the ground-truth labels of all modules are known). At the model application stage, practitioners follow the suggested order of a defect prediction model to inspect/test the modules and stop when the available resources are exhausted. When they stop, they have known the actual labels of the modules they inspect/test: defective or not, which means the TP (number of True Positive modules) and FP (number of False Positive modules) are known. However, the ground-truth labels of the uninspected/untested modules are still unknown, which means the TN (number of True Negative modules) and FN (number of False Negative modules) are unknown. Therefore, SQA practitioners will find conventional performance indicators such as recall, pf, F1, G1, and AUC are not applicable at the model application stage. Different from those conventional performance indicators, ROI and eIFA are applicable at the model application stage as well as at the model evaluation stage.

In the following, for a model $m$, we illustrate how to compute the corresponding ROI and eIFA. Note that, in our context, $m$ can be *model* or ONE. If necessary, we will explicitly mention *model* and ONE.

**ROI** (**R**eturn **O**ver **I**nvest). If practitioners follow a defect prediction model $m$ to inspect/test the predicted defective modules, the invest is the SQA-effort required, and the return is the number of actual faults found (i.e., the number of actual defective modules found in our study). For a target (test) project $p$ containing $k$ modules, we assume that:

- $[r_1, r_2, \ldots, r_k]$: the defect-proneness ranking (in descending order) given by model $m$;
- $n_i$: the actual defect label of the $i$-th module (0 for non-defective and 1 for defective);
- $s_i$: the SLOC of the $i$-th module;
- $x$: the number of the top-ranked modules to be inspected/tested recommended by model $m$.

Consequently, the total number of defective modules in project $p$ is:

$$n = n_1 + n_2 + \cdots + n_k \tag{1}$$

and the total code size in SLOC in project $p$ is:

$$s = s_1 + s_2 + \cdots + s_k \tag{2}$$

From the above information, we can easily conclude that TP (number of True Positive modules), PII (the Percentage of Instances Inspected/tested), and PCI (the Percentage of Code Inspected/tested) are respectively:

$$\text{TP} = \sum_{i=1}^{x} n_i \tag{3}$$

$$\text{PII} = \frac{x}{k} \tag{4}$$

$$\text{PCI} = \frac{1}{s}\sum_{i=1}^{x} s_i \tag{5}$$

As mentioned before, two kinds of SQA-effort are involved during code inspection/testing: code inspection/testing effort and context-switch effort. In order to simplify the calculation, we follow prior studies [46,75] to use PCI as the proxy of code inspection/testing effort and use PII as the proxy of context-switch effort. As a result, the ROI of model $m$ can be defined as:

$$ROI = \frac{TP}{a * PCI + (1-a) * PII} \tag{6}$$

By definition, ROI denotes the number of defective modules found per unit SQA-effort. In practice, it may be challenging to determine the value of parameter $a$. In MATTER, the comparison of model *model* with ONE is conducted under the following SQA-effort-aligned settings: SNM (Same Number of Modules) and SSC (Same Size of Code). Given the above fact, we can simplify the definition of ROI under SNM as:

$$ROI = \frac{TP}{PCI} \tag{7}$$

The reason is that *model* and ONE have the same PII under SNM. In this context, removing PII from the original definition (6) of ROI can avoid the determination of the parameter $a$, thus simplifying their comparison. Similarly, under SSC, the definition of ROI can be simplified as:

$$ROI = \frac{TP}{PII} \tag{8}$$

After the above simplification, ROI denotes the number of actual defective modules found per unit code inspection/testing effort under SNM and denotes the number of actual defective modules found per unit context-switch effort under SSC. By comparing ROI, we are able to determine whether the new defect prediction model *model* is more cost-effective in finding faults than the simple baseline model ONE.

**eIFA** (SQA-**e**ffort required in **I**nitial **F**alse **A**larms). If practitioners follow a defect prediction model $m$ to inspect/test the predicted defective modules, the SQA-effort required in IFA includes the code inspection/testing effort required involved in IFA and the context-switch effort involved in IFA. For a target (test) project $p$ containing $k$ modules, we assume that $y$ is the number of top-ranked modules to be inspected/tested before the first true defective module in the defect-proneness ranking (in descending order) given by model $m$. Consequently, the code inspection/testing effort and the context-switch effort involved in IFA are respectively:

$$\text{PCI}_{\text{IFA}} = \frac{1}{s}\sum_{i=1}^{y} s_i \tag{9}$$

$$\text{PII}_{\text{IFA}} = \frac{y}{k} \tag{10}$$

As a result, the eIFA of model *m* can be defined as:

$$eIFA = a * PII_{IFA} + (1-a)PCI_{IFA} \tag{11}$$

In defect prediction, IFA has been used to evaluate the number of modules that are unnecessarily inspected/tested before finding the first true defective module [42,46,76]. In our study, we use eIFA (with a = 0.5) instead of IFA to evaluate the involved SQA-effort, rather than only the number of the initial false alarms. By comparing eIFA, we are able to determine whether the new defect prediction model *model* requires less SQA-effort in finding the first defective module than the simple baseline model ONE.

In addition to ROI and eIFA, we include a commonly used traditional non-effort-aware indicator, **MCC** (its definition is equation 12), as the third core performance indicator, to provide a complementary view for the prediction performance. Previous systematic literature reviews [35,77] revealed that the most commonly used traditional performance indicators in the defect prediction community include recall, precision, F1, AUC, and MCC. We select MCC rather than other indicators for the following reasons: (1) as discussed by Yao and Shepperd [78], MCC has the advantage that it measures whether the prediction is better than a random model (a MCC larger than 0 means better than a random model); (2) MCC considers every aspect in the confusion matrix (TP, FP, TN, and FN) [79]; (3) accuracy, recall, precision, and F1 are considered to be unstable for imbalanced defect dataset and therefore are not recommended by several studies [79-81]; and (4) AUC is a threshold-independent indicator calculated on the whole rank of modules, which does not depict how well is in the front part of the defectiveness rank that is of practical value for SQA practitioners.

$$MCC = \frac{TP \times TN - FP \times FN}{\sqrt{(TP+FP)(TP+FN)(TN+FP)(TN+FN)}} \tag{12}$$

In particular, under SNM, if *model* has a higher MCC compared with ONE, then we can easily conclude that *model* must have: (1) a higher recall, precision, F1, G1, G2 (The geometric mean of recall and precision), G3 (The geometric mean of recall and 1-PF), and balance [82]; and (2) a lower PF (probability of false alarms). This means that under SNM, MCC is coincident with the other commonly used traditional performance indicators in determining which model is superior. The detailed proof can be found in the online **Appendix A** [83].

In summary, we suggest using the following three core indicators as a starting point for consistent model performance comparison under MATTER: ROI, eIFA, and MCC. The former two indicators are effort-aware, which can shed light on whether a new defect prediction model *model* has a higher practical value than ONE when SQA-effort is considered. The last indicator provides a complementary view of the prediction performance for those practitioners that care about the (traditional) non-effort prediction performance. It should be pointed out that ROI, eIFA, and MCC are just three core indicators we recommended to evaluate the real progress of defective prediction models. If needed, such a set of indicators can be expanded with other performance indicators for specific situations. We report detailed prediction results of models in this paper and give an implementation of a set of commonly used performance indicators in our replication kit to support such expansion. The key is to keep the ongoing use of the same appropriate performance indicators when evaluating and comparing model performance within studies and across studies.

## 5 EXPERIMENTAL SETUP

In this section, we introduce the details of the experimental setup used to evaluate the effectiveness of our framework MATTER, including the research questions (Section 5.1), compared defect prediction models (Section 5.2), studied datasets (Section 5.3), and evaluation strategy (Section 5.4).

### 5.1 Research questions

Our study set ups four research questions. The former three RQs focuses on the evaluation of our framework MATTER itself, i.e., investigate the necessity of the SQA-aligned threshold setting (RQ1), the appropriateness of using ROI, eIFA, and MCC as the core indicators (RQ2), and the advantage of using ONE as the baseline model (RQ3). The last RQ focuses on the investigation of the status quo of defect prediction by MATTER.

**RQ1:** Is it necessary to align the SQA-effort in MATTER when comparing model prediction performance?

The purpose of RQ1 is to investigate the importance of the SQA-aligned threshold setting in model evaluation. To this end, we conduct a comparison experiment using eight typical defect prediction models (see details in Section 5.2) due to the fact that: (1) they have been shown a promising defect prediction performance in previous studies; (2) they have been recommended as the baseline models in the literature or are potential to be the baseline models in defect prediction and hence may have an important influence on subsequent defect prediction studies; and (3) their required SQA-effort under default (i.e., effort-unaligned) thresholds exhibits a large variance. In previous studies, many of them have been compared in terms of prediction performance under the unaligned SQA-effort. This provides a good opportunity for us to investigate the impact of the SQA-effort-aligned threshold setting on model comparison. It is worth noting that, SNM and SSC, two threshold-aligned settings in MATTER, are meaningful in real development scenario. In practice, given the defect-proneness ranking lists at the module level produced by different models, users naturally want to know which one is better in prediction when selecting the same number of modules (SNM) or the same size of code (SSC) for inspection/testing. If the performance ranking among prediction models has a large change before and after the SQA-effort-alignment, this will indicate that the unaligned-threshold-setting distorts the conclusion of the model comparison in prediction performance (from the viewpoint of a fair comparison). In this case, in order to get an objective performance ranking between prediction models, it is necessary to align the SQA-effort in MATTER when comparing model prediction performance. The answer of RQ1 will help us to determine the necessity of the SQA-aligned threshold setting.

**RQ2:** Are the three indicators in MATTER appropriate for evaluating defect prediction models?

The purpose of RQ2 is to investigate whether ROI, eIFA, and MCC are appropriate to be used as the performance indicators in MATTER. In Section 4.3, we highlight the fact that: (1) ROI and eIFA incorporate the SQA-effort and hence are more informative than the counterparts when evaluating the cost-effectiveness of a defect prediction model; and (2) due to the chance-corrected and unbiased characteristics, MCC is more informative than the counterparts when evaluating the classification performance of a defect prediction model. Given the above theoretical advantages, it appears to be a reasonable choice to use them as the performance indicators when comparing different defect prediction models. However, another perspective still needs to be explored: whether they are proxies of the counterpart indicators when producing the performance ranking of different models over multiple data sets? In the literature, a lot of counterpart indicators are available. For each (say *X*) of the three chosen indicators, if it produces a different conclusion in terms of the ranking of different prediction models when compared with the counterparts, this means that *X* is not a proxy of the counterparts from the viewpoint of performance comparison. In other words, X plays a role in producing performance ranking that cannot be replaced by the counterparts. In this case, combining the theoretical advantage (more informative), the above result can better justify the choice of *X* as a performance indicator in MATTER. In this way, we can empirically examine whether it is a reasonable choice to use ROI, eIFA, and MCC as the indicators in MATTER. In our study, given a performance indicator, we apply the non-parametric Scott-Knott ESD test to produce the performance ranking of different models over multiple data sets (see Section 5.4 for detail). For ROI, the considered counterpart indicator is recall@X% (the most common value of X% is 20% SLOC, also named as PofB20 [75] or CostEffort@20% [46]), a commonly used effort-aware indicator [75]. For eIFA, the considered counterpart is IFA, a frequently used indicator in recent studies [5,46,76,84]. For MCC, the considered counterparts are recall, precision, F1, G1, and PF. In particular, we employ the same eight defect prediction models used in RQ1 to investigate RQ2.

**RQ3:** Is ONE a good baseline defect prediction model used in MATTER?

The purpose of RQ3 is to investigate whether ONE is a strong baseline model in MATTER by comparing defect prediction models that have been recommended as the baseline defect prediction models in the literature or are potential to be baseline models (see details in section 5.2). As highlighted in Section 4, to be a global baseline model, a model should satisfy "3S properties". Since ONE is only based on the SLOC feature in a test set, ONE is simple and stable. The remaining problem needed to be answered is: is ONE strong enough in prediction ability? Therefore, RQ3 investigates whether ONE is strong enough in defect prediction compared with those models that are recommended to be or have the potential to be the baseline models in the literature. Specifically, we employ the same eight defect prediction models used in RQ1 as the compared models to investigate RQ3. In order to obtain the performance ranking of different models, we apply the non-parametric Scott-Knott ESD test over multiple data sets (see Section 5.4 for detail). If the experimental result shows that ONE has a high performance ranking, this means that ONE is a good baseline model in MATTER to provide "the global reference point" for evaluating newly proposed software defect prediction models.

**RQ4:** How far have we really come in defect prediction according to MATTER?

The purpose of RQ4 is to investigate the status quo of defect prediction by evaluating the practical effectiveness of the representative supervised defect prediction models under MATTER. Specifically, in RQ4, the representative defect prediction models in the current SDP field will be evaluated under MATTER. To avoid implementation bias, we select representative models whose implementations are publicly available (see detail in Section 5.2). This RQ would shed light on the necessity to evaluate defect prediction models under a consistent model performance evaluation framework to get the real progress in practical effectiveness. Clearly, the experimental results obtained from RQ4 can reveal whether the existing studies have made outstanding progress in defect prediction.

## 5.2 Compared defect prediction models

The experiment to answer the first three RQs is conducted on eight defect prediction models that are recommended to be or have the potential to be the baseline models. In the literature, a number of models, such as Bellwether [11], EASC_E, EASC_NE [46], ManualDown [60-62], and ManualUp [60-62] have been shown a good performance and have been explicitly suggested as the baseline models. In addition, a number of unsupervised models such as SC [34], CLA [33], and FCM [85], which are not very complex, do not need any training data, and have been proved to perform well in previous comparative experiments, have the potential to be the baseline models.

- *Bellwether.* Given *N* projects from a community, there is an exemplary project (called bellwether) whose data yields the best predictions on all others. According to [11], we should use the bellwether project data to train a model for predicting defects in new projects in the community. In practice, the defect prediction on a target project proceeds as follows. First, take the target project as the holdout and select the best-performing project from the remaining projects. The selection is conducted in a round-robin style, and G1 is the default performance indicator used to select the "bellwether". Second, build a defect prediction model with the bellwether project data (the default modeling

- *technique is random forest). Third, apply the resulting model to predict defects for the modules in the target project. In our study, we use the replication package shared by Krishna et al. [11] to build the Bellwether model.

- *EASC*. According to [46], EASC indeed consists of two models: *EASC_E* and *EASC_NE*. The former is suggested as the baseline model under the effort-aware performance evaluation, while the latter is suggested as the baseline model under the non-effort-aware performance evaluation. For a target project, both *EASC_E* and *EASC_NE* first rank the predicted defective modules and non-defective modules separately and then concatenate them to obtain a final ranking. The difference is that *EASC_E* ranks the modules in descending order by *score*/LOC, while *EASC_NE* ranks the modules in descending order by *score*×LOC. Here, LOC denotes the module size as measured by lines of code, while *score* is the defect-proneness predicted by a classier. In default, the used classifier is Naïve Bayes. In our study, we strictly follow the process described in Ni et al.'s paper [46] to build the *EASC_E* and *EASC_NE* models.

- *SC.* Zhang et al. [34] used a connectivity-based classifier, spectral clustering (SC) [86], for defect prediction. SC is performed on a graph consisting of nodes and edges. Each node represents a module, and each edge represents the connection between modules, and the weight is measured by the similarity of metric values between two modules. To construct an unsupervised defect prediction model, SC first minimizes the normalized cut of the graph. In this way, SC can achieve a low similarity across the two subgraphs and gain high similarity within each subgraph. Then, SC labels each of the two subgraphs. Based on the insight that for most metrics, the defective modules generally have larger values than the clean modules, the subgraph that has a larger average sum of metric values is labeled as "defective", and the other subgraph is labeled as "clean". In our study, we use the replication package shared by Zhang et al. [34] to build the SC model.

- *CLA.* The key idea of CLA is to label an unlabeled dataset by using the magnitude of metric values [33]. Specifically, CLA believes that the defective modules generally have larger values than the clean modules for most metrics. CLA includes two steps: **C**lustering instances and **LA**beling instances in clusters. First, count for each module how many of their metrics exceed the median value of a metric in the target project (referred as K). Then, group the modules with the same K value into the same cluster. Third, label the top half clusters as defective and the bottom half clusters as clean. In our study, we use the replication package shared by Nam et al. [33] to build the CLA model.

- *FCM.* Fuzzy C-Means (FCM) is a widely used fuzzy clustering algorithm [87] and was found to be one of the best unsupervised techniques in [85]. FCM is a clustering algorithm in which each data point belongs to each cluster center on the basis of the distance between the center and itself to a certain degree. To get prediction results of *n* modules in the target project using FCM, first, apply the FCM algorithm with clustering as 2 to get a degree score matrix with *n* rows and 2 columns. Each column indicates the degree that a module belongs to a cluster. Then, to label each module, the module is considered to belong to the cluster in which its degree is bigger than in another cluster. At last, the cluster that contains more instances is considered as "clean", and another cluster is considered as "defective". In our study, we strictly follow the process described in [85] to build the FCM model.

- *ManualDown and ManualUp.* ManualDown and ManualUp [60-62] are two module size-based unsupervised baseline models recommended in prior studies [9]. Given a target project, ManualDown and ManualUp respectively predict the defect proneness of a module *m* as *score*(*m*) = SLOC(m) and *score*(*m*) = 1/SLOC(*m*). The intuition behind ManualDown is that larger modules may have more defects and therefore should be inspected/tested first. In contrast, the intuition behind ManualUp is that smaller modules may have larger defect density and therefore should be inspected/tested first.

In RQ4, we chose three representative CPDP models and one state-of-the-art HDP model. CamargoCruz09-NB, Amasaki15-NB, and Peters15-NB are the top three best-performing models in a recent large-scale comparative study for cross-project defect prediction (CPDP) models [88], while KSETE [5] is the state-of-the-art heterogeneous defect prediction (HDP) models.

- *CamargoCruz09-NB, Amasaki15-NB, and Peters15-NB*. Herbold et al. [88] conducted a comparative analysis of 114 CPDP models on 86 defect data sets. They reported that the three best-performing models were CamargoCruz09-NB, Amasaki15-NB, and Peters15-NB[1]. CamargoCruz09-NB standardizes the target and training data using the logarithm and the target data as a reference point. Amasaki15-NB combines attribute selection and relevancy filtering: first, the data is log-transformed; then, the metrics that are closest to any other metric are moved out from the whole dataset; finally, the instances that are the closest to any other instance are filtered out. Peters15-NB leveraging LACE2 as an extension of CLIFF and MORPH for privatization. All of those three models use Naïve Bayes as the classifier. In our study, we use the replication package shared by Herbold et al. [88,89] to build these three models.

- **KSETE**. Tong et al. [5] proposed a kernel spectral embedding transfer ensemble (KSETE) model for HDP. In nature, this model combines kernel spectral embedding, transfer learning, and ensemble learning to find the latent common feature space for the source and target datasets. First, use SMOTE on the source dataset to alleviate the class-imbalance problem and normalize the source dataset and the target dataset with Z-score. Second, randomly sample the same number of modules from the source and target datasets. Third, perform a kernel spectral embedding to get the projected source and target datasets. The goal is to find a latent common kernel feature subspace that the subspace not only maximizes the similarity between the projected source and target datasets but also preserves the intrinsic characteristic of both the source and target datasets. The process is repeated to find a series of pairs of projected source and target datasets. Finally, build a classifier on each projected source dataset and combine multiple classifiers to predict the labels of the projected target dataset using transfer ensemble learning. In our study, we use the KSETE replication package shared by Tong et al. [5] to build the KSETE model.

Note that all the investigated models in our experiment are evaluated on the same target data sets. Therefore, they are not only comparable with ONE but also comparable with each other. We had planned to include top-core [47] and MSMDA [2] when investigating RQ4, as their replication kits are available. However, for top-core, the predicted defectiveness of a module depends on the "coreness" computed from a graph extracted from a target project's source code. In [47], Qu et al. provided such information for only 8 test projects. For the other test projects used in our study, such information is unavailable. MSMDA is a multi-source selection based manifold discriminant alignment model, which has a high memory requirement. Our experiment shows that MSMDA does not scale to a large number of source projects, as "out-of-memory failure" is reported. Considering the above facts, we do not include top-core and MSMDA when investigating RQ4. Nonetheless, in our online Appendix B [83], we report: (1) the comparison of ONE with top-core using the same 8 projects as used in [47]; and (2) the comparison of ONE with MSMDA using the same 28 projects as used in [2]. The key message is that top-core underperforms ONE and MSMDA does not lead to a practically important progress.

### 5.3 Studied dataset

Table 4 summarizes the statistical information of the datasets used in our study. The first and the second columns report the name of the dataset and the name of each project, respectively. The third column reports the number of versions a project contains. The fourth column reports the number of instances (for projects that contain one release) or the range of the number of instances (for projects that contain more than one release). The fifth column reports the percentage of defective modules or the range of percentage of modules. The last two column reports the number of metrics and the types of metrics, respectively.

---

[1] Note that their original paper [88] reported that CamargoCruz09−DT, Turhan09−DT, and Menzies11−RF were the three best performing models. In [47], Ni et al. used them as the compared models to demonstrate the effectiveness of their proposed EASC. However, as immediately corrected by Herbold et al. [89], the three best-performing cross-project defect prediction models were CamargoCruz09-NB, Amasaki15-NB, and Peters15-NB. Therefore, in our study, we use CamargoCruz09-NB, Amasaki15-NB, and Peters15-NB as the compared models for ONE.

Table 4. An overview of the studied datasets

| Dataset | Project | #Versions | #Instances | %Defective | #Metrics | Metric type |
|---|---|---|---|---|---|---|
| AEEEM | eclipse | 1 | 997 | 21% | 31 | product, process |
| | equinox | 1 | 324 | 40% | | |
| | lucene | 1 | 691 | 9% | | |
| | mylyn | 1 | 1862 | 13% | | |
| | pde | 1 | 1497 | 14% | | |
| JURECZKO | ant | 5 | 125~745 | 11%~%26 | 20 | product |
| | arc | 1 | 234 | 12% | | |
| | berek | 1 | 43 | 37% | | |
| | camel | 4 | 339~965 | 4%~%36 | | |
| | ckjm | 1 | 10 | 50% | | |
| | e-learning | 1 | 64 | 8% | | |
| | forrest | 1 | 29 | 17% | | |
| | ivy | 3 | 111~352 | 7%~%57 | | |
| | jedit | 5 | 272~492 | 2%~%33 | | |
| | kalkulator | 1 | 27 | 22% | | |
| | log4j | 3 | 109~205 | 25%~%92 | | |
| | lucene | 3 | 195~340 | 47%~%60 | | |
| | nieruchomosci | 1 | 27 | 37% | | |
| | pbeans | 2 | 26~51 | 20%~%77 | | |
| | pdftranslator | 1 | 33 | 45% | | |
| | poi | 4 | 237~442 | 12%~%64 | | |
| | redaktor | 1 | 176 | 15% | | |
| | serapion | 1 | 45 | 20% | | |
| | skarbonka | 1 | 45 | 20% | | |
| | sklebadg | 1 | 20 | 60% | | |
| | synapse | 3 | 157~256 | 10%~%34 | | |
| | systemdata | 1 | 65 | 14% | | |
| | szybkafucha | 1 | 25 | 56% | | |
| | termoproject | 1 | 42 | 31% | | |
| | tomcat | 1 | 858 | 9% | | |
| | velocity | 3 | 196~229 | 34%~%75 | | |
| | workflow | 1 | 39 | 51% | | |
| | wspomaganiepi | 1 | 18 | 67% | | |
| | xalan | 4 | 723~909 | 15%~%99 | | |
| | xerces | 4 | 162~588 | 15%~%74 | | |
| | zuzel | 1 | 29 | 45% | | |
| ReLink | Apache | 1 | 194 | 51% | 40 | product |
| | Safe | 1 | 56 | 39% | | |
| | Zxing | 1 | 399 | 30% | | |
| MA-SZZ-2020 | flume | 10 | 272~574 | 4%~%32 | 44 | product |
| | mahout | 10 | 1027~1267 | 17%~%29 | | |
| | maven | 10 | 318~703 | 4%~%13 | | |
| | shiro | 10 | 381~493 | 3%~%7 | | |
| | zeppelin | 10 | 129~413 | 21%~%38 | | |
| IND-JLMIV+R | 23 projects | 59 | 103~1415 | 5%~20% | 22 | process |

As can be seen in Table 4, we use AEEEM [90], JURECZKO [91], ReLink [92], MA-SZZ-2020 [54], and IND-JLMIV+R [93] as the datasets to conduct our experiment. AEEEM, JURECZKO, and ReLink are three defect datasets that are frequently used in the field of defect prediction. MA-SZZ-2020 [54] and IND-JLMIV+R [93] are two new quality defect datasets published in 2020. The original IND-JLMIV+R data set consists of 395 versions, many of which contain few defective modules or even no defective modules. In order to ensure a reasonable analysis for defect prediction, we filter IND-JLMIV+R by the following requirements: each release has at least 100 instances, at least 5% of the instances are defective, and there are at least 10 defective instances. As a result, the resulting IND-JLMIV+R data sets consists of 59 releases of 23 projects. In total, our dataset contains 179 versions of 67 projects.

### 5.4 Evaluation Strategy

**Training-test data split**. We apply the all-to-one strict cross-project defect prediction (CPDP) setting for the studied supervised CPDP models to split the training and test sets. Specifically, for each release in a project as the test set, we use all releases in external projects within the same dataset as a whole as the training set. The only exception is KSETE, for which we apply a one-to-one CPDP setting. The reason is that KSETE depends on matrix computations, which does not scale to a large training set. For each release in a project as the test set, we use each release in external projects within the dataset as a training set. For each test set, we report the mean performance across all training sets for KSETE. All models in our experiments are compared on the same test sets.

**Performance indicators**. We use three core performance indicators in MATTER (MCC, ROI, and eIFA) in our experiments. Note that: (1) the higher the MCC and ROI, the better the prediction model is; and (2) the lower the eIFA, the better the prediction model is. As stated in Section 4.3, such a set of indicators can be expanded with performance indicators for specific situations if needed, and such expansion is supported by our replication kit [94].

**Statistical comparison**. We use the non-parametric Scott-Knott ESD test [5,58,95] to divide defect prediction models into different groups. The Scott-Knott ESD test is currently one well-accepted statistical test for comparing multiple defect prediction models over multiple data sets [2,95-98]. In our experiment, we follow [95] to first rank models in terms of prediction performance on each test set to get a ranking list (the purpose is to control for dataset-specific model performance, as some datasets may be more or less susceptible to bias than others) and then use the ranking list as the input to apply the non-parametric Scott-Knott ESD test. The non-parametric Scott-Knott ESD test produces groups of defect prediction models and ensures that the magnitude of models' ranking difference within the same group is negligible, and the magnitude of models' ranking difference between groups with different rankings is non-negligible. The Cliff $\delta$ effect size [99,100] is used to decide the magnitude of the ranking differences between compared models. The magnitude threshold is: negligible for $|\delta| < 0.147$, small for $0.147 <= |\delta| < 0.33$, medium for $0.33 <= |\delta| < 0.474$, and large for $|\delta| >= 0.474$ [101]. In our experiment, we use the implementation of the non-parametric Scott-Knott ESD test in the ScottKnottESD R package (available at https://github.com/klainfo/ScottKnottESD 3.0 version).

## 6 EXPERIMENTAL RESULTS

In this section, we conduct a large experiment based on the above setup and analyze the experiment results to answer the corresponding research questions.

### 6.1 RQ1: Is it necessary to align the SQA-effort in MATTER when comparing prediction performance?

Fig. 7 reports the distributions of the required code inspection effort (in terms of PCI) and context switch effort (in terms of PII) over the test sets for eight defect prediction models under: (a) default thresholds for individual models, (b) SNM (PII=20%), and (c) SSC (PCI=20%). Note that, under SNM (PII=20%), due to the fact that 20% × *total_number_of_modules* may not be an integer, the PII on some test sets cannot achieve exactly 20% (as close as 20%, not exceed 20%). Furthermore, in practice, given a defect prediction model at the module granularity, the minimum unit for selecting the code to be inspected is one module (rather than one line of code). As a result, under SSC (PCI=20%), the PCI on some test sets cannot achieve exactly 20% (as close as 20%, not exceed 20%). In order to get a clearer picture of the difference in SQA-effort under unaligned and aligned thresholds, Fig. 8 uses boxplots to summarize the distribution of the median SQA-effort for the eight investigated models.

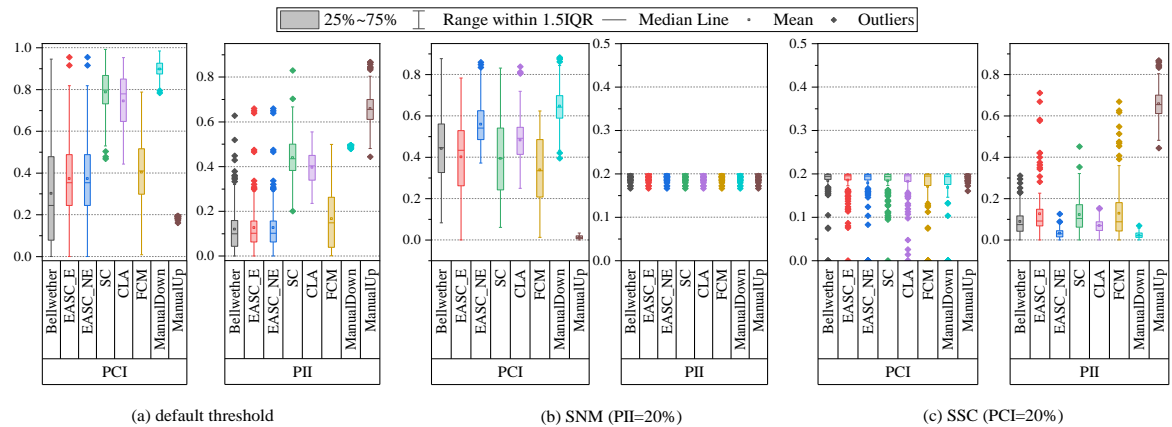

Figure 7: The distribution of the required SQA-effort over the test sets for eight investigated models under: (a) default thresholds, (b) SNM (PII=20%), and (c) SSC (PCI=20%)

From Fig. 7(a), we can see that, under individual default thresholds, both the code inspection effort and the context switch effort required by defect prediction models are quite different from each other. The median PCI required by models ranges from about 20% to 90% and the median PII required by models ranges from about 10% to 65%. However, from Fig. 7(b) and Fig. 7(c), we can see that the variance of the required SQA-effort has been considerably reduced under both SNM and SSC. The above observation is clearly confirmed in Fig. 8. The key message from Fig. 7 and Fig. 8 is that, without the alignment of thresholds, there is a very large variance in SQA-effort over different defect prediction models. As a result, it is not possible to fairly compare their prediction performance. The reason is that we would not know whether a better performance should be attributed to the prediction power of a defect prediction model or the influence of the unaligned SQA-effort. After the thresholds are aligned, different defect prediction models have a relatively comparable SQA-effort, which would promote a fair performance comparison.

What if we do not align the thresholds when comparing different defect prediction models? We next use MCC and PF as an example to exhibit the influence of unaligned thresholds on the performance ranking of different models (the observations from other performance indicators are similar). Fig. 9 reports the heatmaps of the magnitude of ranking difference of every two paired models and the grouping results according to the non-parametric Scott-Knott ESD test under (a) default thresholds, (b) SNM (PII=20%), and (c) SSC (PCI=20%). For example, in Fig. 9(a), when models are compared under their default thresholds in terms of MCC on the test sets, we can see that: (1) CLA, ManualDown, and SC are grouped in the first (best) rank, (2) the difference among CLA, ManualDown, and SC are negligible, (3) CLA, ManualDown, and SC perform better than EASC_E, EASC_NE, FCM, and Bellwether in a medium effect size, and (4) CLA, ManualDown, and SC perform better than ManualUp in a large effect size.

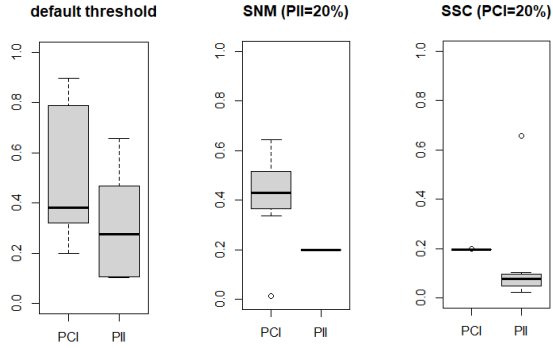

Figure 8: The distribution of the median SQA-effort over eight investigated models under: (a) default thresholds, (b) SNM (PII=20%), and (c) SSC (PCI=20%)

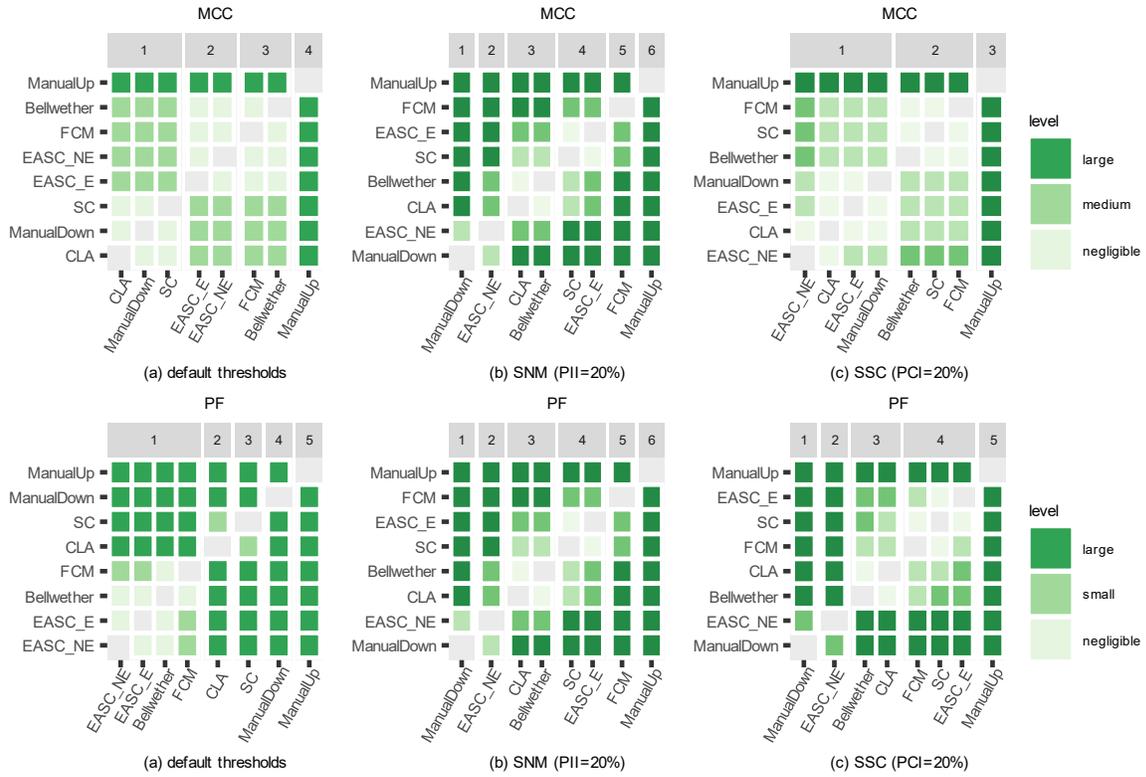

Figure 9: Heatmap of the magnitude of ranking difference of every two paired models of MCC (PF) and the grouping results according to the non-parametric Scott-Knott ESD test under: (a) default thresholds, (b) SNM (PII=20%), and (c) SSC (PCI=20%). The darker the color, the larger the magnitude of the ranking difference. The more to the left and the more on the bottom of a model, the better its MCC (PF) ranking distribution.

From Fig. 9, we can see that the threshold-alignment setting may lead to a dynamic change of the rank of a defect prediction model.

- *A defect prediction model may have a top rank under the default threshold setting but have a low rank under the threshold-alignment setting.* Consider SC as an example. In terms of MCC, according to the non-parametric Scott-Knott ESD test, SC has a rank of 1 (the best rank) under the default threshold setting (i.e., the threshold-unalignment setting). However, its rank drops to 4 under SNM and drops to 2 under SSC. When looking at Fig. 7(a), we can see that SC incurs very high PCI and PII under the default threshold setting compared with the other models, i.e., SC suggests considerably more code and modules to be inspected. However, after aligning the thresholds, SC suggests comparable code and modules to be inspected. This indicates that, under the default threshold setting, the SQA-effort incurred by SC has a large contribution to the high MCC rank of SC. As can be seen, if we do not control the influence of the SQA-effort when comparing SC with the other models, SC will have a spurious high MCC rank that falsely accentuates its real effectiveness in defect prediction.
- *A defect prediction model may have a low rank under the default threshold setting but have a top rank under the threshold-alignment setting.* Consider ManualDown as an example. In terms of PF, according to the non-parametric Scott-Knott ESD test, ManualDown has rank of 4 (the second to last rank) under the default threshold setting (in [46], Ni et al. reported that ManualDown was much inferior to EASC_N in terms of PF). However, its rank rises to 1, regardless of whether the SNM or SSC setting is considered. When looking at Fig. 7(a), we can see that, compared with the other models, ManualDown suggests many more modules to be inspected and thus results in a high PF. Clearly, this is an unfair comparison setting for ManualDown. As shown in Fig. 7(b) and Fig. 7(c), after aligning the threshold, ManualDown suggests a comparable number of modules and a comparable amount of code for inspection compared with the other models. The threshold-alignment setting leads to a dynamic change of the FP rank of ManualDown (the best rank). Indeed, the performance rankings under the default threshold setting and SNM has a Spearman correlation of only 0.116, while the rankings under the default threshold setting and SSC has a Spearman correlation of only 0.163. As can be seen, if we do not control the influence of the SQA-effort when comparing ManualDown with the other models, ManualDown will have a spurious low PF rank that falsely obscures its real effectiveness in defect prediction.

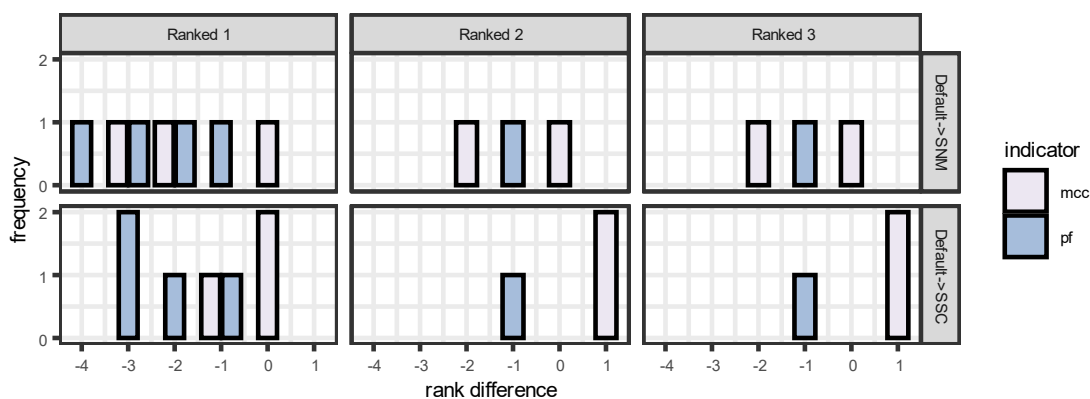

Figure 10: The difference in the ranks for the metrics according to the SK ESD test grouping result. The bars indicate the frequency of the rank difference between the comparison under default thresholds and the comparison under SQA-aligned thresholds. In each column, models are grouped in Rank 1, Rank2, or Rank 3 in the comparison under default threshold. For each row, we investigate how ranks change from the default threshold setting to the SQA-effort threshold aligned settings (Default->SNM or Default->SSC).

In order to estimate the influence that the threshold-alignment setting has on the performance ranking, we compute the difference in the ranks of the models that appear in the top-three ranks under the threshold-unalignment setting. For example, if *m* has a rank of 1 under the threshold-unalignment setting and has a rank of 4 under the threshold-alignment setting, then the rank difference of *m* would be 1–4 = –3. Fig. 10 reports the rank differences for the models according to their prediction performance. As can be seen, the models in the top 3 ranks are in general unstable. In other words, whether aligning the thresholds may have a large influence on the resulting ranking of different defect prediction models. Combining the above observations with the SQA-effort shown in Fig. 7, we can conclude that, if the thresholds are not aligned, the SQA-effort exhibits a large variance, which can have a strong confounding effect on the performance ranking of different prediction models. Clearly, the threshold-alignment setting can mitigate the confounding effect and hence leads to a fair performance comparison. In particular, after the thresholds are aligned, only code inspection effort (under SNM) or context switch effort (under SSC) is needed to be considered in the computation of ROI, which facilitates the accurate comparison of the cost-effectiveness of defect prediction models. In this sense, there is a strong need to align the SQA-effort in MATTER when comparing prediction performance. Otherwise, the performance ranking of different defect prediction models will be distorted, which may result in misleading conclusions.

> ***Conclusion.*** *Under the unaligned threshold setting, the required SQA-effort varies considerably with defect prediction models, which may have a strong confounding effect on performance evaluation. The threshold-alignment setting can mitigate the confounding effect and hence leads to a fair performance comparison. Therefore, it is necessary to use SNM and SSC to align the SQA-effort in MATTER when comparing prediction performance. This is especially true if we also consider the fact that SNM and SSC are two important and meaningful scenarios in defect prediction.*

**6.2 RQ2: Are the three indicators in MATTER appropriate for evaluating defect prediction models?**

Table 5 reports the Spearman's [102] and Kendall's [103] correlations of ROI with recall@20%, PII, and PCI. The purpose is to observe whether ROI produces a different ranking of defect prediction models when compared with the counterparts. Note that, recall is exactly recall@20% (20% stands for 20% of total modules) under SNM (PII=20%), while recall is recall@20% (20% stands for 20% of total SLOC) under SSC (PCI =20%). In particular, we first apply the non-parametric Scott-Knott ESD test to produce the rankings of different models over multiple data sets in terms of ROI, recall@20%, PII, and PCI, separately. Then, we compute the correlation between two resulting model rankings. According to the definition of Spearman's and Kendall's correlations, the closer the absolute value of the correlation to 1, the stronger the correlation between the two model rankings (a correlation larger than 0 indicates a positive correlation while a correlation smaller than 0 indicates a negative correlation). For example, under SSC (PCI=20%), according to the non-parametric Scott-Knott ESD test, EASC_NE, ManualDown, CLA, Bellwether, EASC_E, SC, FCM, and ManualUp have a ranking of (1, 1, 2, 2, 2, 2, 2, 3) in terms of ROI and have a ranking of (5, 6, 3, 4, 2, 3, 3, 1) in terms of recall@20% (see more detail ranking results in [94]). Consequently, the Spearman's correlation between those two rankings is −0.887, indicating a strong negative correlation between ROI and recall@20% under SSC.

Table 5. The Spearman's and Kendall's correlations of ROI with recall@20% (also called PofB20 or CostEffort@20%), PII, and PCI for eight compared defect prediction models.

|          | SNM (PII=20%) | | SSC (PCI=20%) | |
|----------|---------------|--------|---------------|--------|
|          | Recall | PCI | Recall | PII |
| Spearman | −0.128 | −0.134 | −0.887 | −0.893 |
| Kendall  | −0.113 | −0.123 | −0.825 | −0.842 |

From Table 5, we can see that ROI produces a different performance ranking compared with the rankings produced by the counterpart indicators, regardless of whether SNM (PII=20%) or SSC (PCI=20%) is considered. More specifically, under SNM (PII=20%), both recall@20% and PCI have weak negative correlations with ROI; under SSC, both recall@20% and PII have strong negative correlations with ROI. Overall, the above results indicate that ROI is not simply a proxy of recall@X%, PCI, or PII. Since ROI produces a different performance ranking and its determination relies on more information (both defects found and SQA-effort cost) that SQA practitioners care about, ROI is more appropriate for evaluating defect prediction models than recall@x%, PCI, or PII.

Table 6 reports the correlation of eIFA with IFA, $PCI_{IFA}$, and $PII_{IFA}$. We can see that the Spearman's and Kendall's correlations range from 0.770 to 0.864, which means eIFA is not totally consistently with its counterpart indicators in determining which models are better in multiple models' comparison (although a strong positive correlation exists). By the definition of eIFA, the reason should be attributed to the extra information provided by the combination of $PCI_{IFA}$ and $PII_{IFA}$, which leads to different model comparison results compared with the counterpart indicators. Therefore, we prefer eIFA for evaluating defect prediction models in MATTER.

Table 6. The Spearman's and Kendall's correlations of eIFA with IFA, $PCI_{IFA}$, and $PII_{IFA}$ for eight compared defect prediction models.

|  | IFA | $PCI_{IFA}$ | $PII_{IFA}$ |
|---|---|---|---|
| Spearman | 0.864 | 0.802 | 0.864 |
| Kendall | 0.831 | 0.770 | 0.831 |

Similar to Table 5 and 6, Table 7 reports the correlations of MCC with the other commonly used classification performance indicators. We can see that, under SNM, MCC has a perfect correlation with all the other indicators. The reason is that, on each single test set, MCC is coincident with all the other indicators in determining which model is superior (see the proof in Appendix A). Figure 11 visualizes the relationship of MCC with recall, precision, F1, G1, and PF on several sampled test sets. As can be seen, the rank of a model under MCC is the same as that under recall, precision, F1, G1, and PF. Under SSC, MCC exhibits different comparison results with those indicators: (1) MCC has a strong positive correlation with precision and PF on which model is superior (2) MCC and recall tend to draw opposite conclusions in determining which model performs better, and (3) the correlations between MCC and F1 and G1 are weak. In summary, under SNM, MCC is coincident with all the other indicators in determining the performance ranking; under SSC, MCC tends to produce a different performance ranking. Combining the chance-corrected and unbiased characteristics, we hence select MCC to evaluate defect prediction models in MATTER.

Table 7. The Spearman's and Kendall's correlations of MCC with the other performance indicators for eight compared defect prediction models

|  | SNM (PII=20%) | | | | | SSC (PCI=20%) | | | | |
|---|---|---|---|---|---|---|---|---|---|---|
|  | Precision | Recall | F1 | G1 | PF | Precision | Recall | F1 | G1 | PF |
| Spearman | 1 | 1 | 1 | 1 | 1 | 0.738 | –0.468 | 0.311 | –0.239 | 0.706 |
| Kendall | 1 | 1 | 1 | 1 | 1 | 0.723 | –0.413 | 0.300 | –0.216 | 0.656 |

It is worth noting that MCC, ROI, and eIFA are the three core indicators we recommended to evaluate the real progress of defective prediction models in MATTER. If needed, such a set of indicators can be expanded with performance indicators for specific situations. As stated in [104], preferably, we "should provide all the confusion matrices so that a wide range of metrics can potentially be computed as secondary analysis, in the event that researchers have a preference for a particular, but alternative, metric." Anyway, the key is to keep the ongoing use of the same performance indicators when evaluating and comparing model performance within studies and across studies.

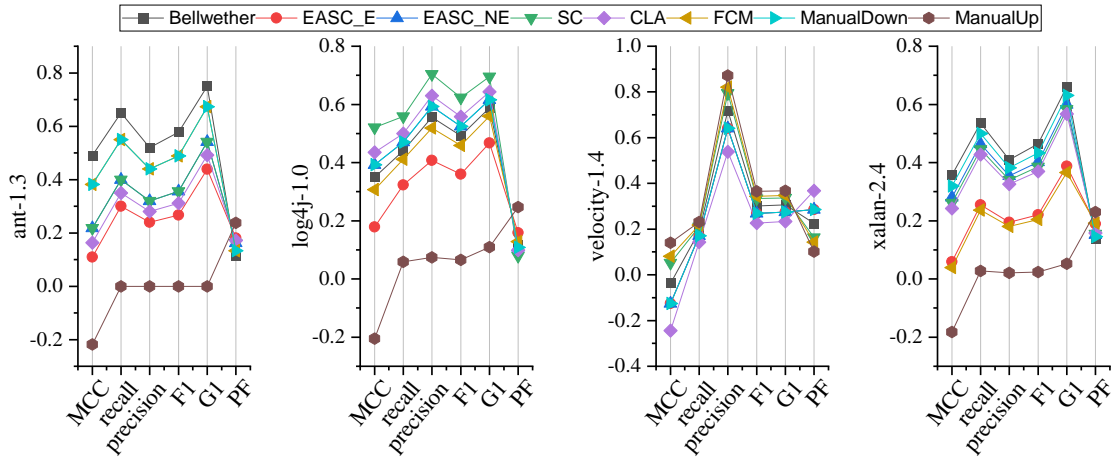

Figure 11: Monotonicity of MCC, recall, precision, F1, G1, and PF among eight compared defect prediction models on the target sets under SNM

> ***Conclusion.*** *ROI, eIFA, and MCC depict important information that SQA-practitioners care about. On the one hand, they are more informative than the counterpart indicators. On the other hand, they are in general not proxies of the counterparts in determining the ranking of different defect prediction models. In this sense, it is appropriate to use ROI, eIFA, and MCC as the core indicators for evaluating defect prediction models in MATTER.*

### 6.3 RQ3: Is ONE a good baseline defect prediction model used in MATTER?

Fig. 12 reports the grouped performance ranking distribution of eight baseline defect prediction models and ONE by the non-parametric Scott-Knott ESD test (the smaller the group number on the top of the plot, the better the performance is). Note that eIFA keeps the same under SNM and SSC, since it is independent of the SQA-effort threshold. Table 8 summarizes the median and mean values of three core indicators under SNM and SSC. The best performance value of each indicator is shown in bold while the worst performance value is shown in underlined. The detailed prediction result for each test set is provided online [94]. The comparison results on individual groups of test sets (i.e., AEEEM, JURECZKO, ReLink, MA-SZZ-2020, and IND-JLMIV+R) can be found in the online Appendix C [83], from which a similar conclusion can be observed compared with the conclusion drawn on all the datasets.

From Fig. 12 and Table 8, we have the following observations.

- *Among the nine compared models, ONE performs from the top to upper-middle level under SNM (PII=20%) and SSC (PCI=20%)*. Under SNM (PII=20%), ONE performs at the top level (as it belongs to group 1) in terms of ROI and eIFA, and at the upper-middle level (as it belongs to group 3 out of 7 groups) in terms of MCC. Under SSC (PCI=20%), ONE performs at the top level (as it belongs to group 1) in terms of MCC and eIFA, and at the second level (as it belongs to group 2 out of 4 groups) in terms of ROI.

- *EASC_NE exhibits a performance similar to ONE in most situations*. Under SNM, ONE achieves a median MCC of 0.219, which is worse than the median MCC of 0.251 for EASC_NE. In terms of MCC, ONE achieves a median value of 0.145 under SSC (PCI=20%), which is close to EASE_NE (a median value of 0.151). According to the rank of ROI (see Fig. 12), ONE is better under SNM but worse under SSC when compared with EASC_NE. The median and mean eIFA values of ONE and EASC_NE are similar.

- *ManualDown and ManualUp perform worse than ONE when both SNM and SSC are considered.* While performing well under SSC, ManualDown performs the worst in terms of ROI under SNM (the reason is that it requires a lot of code inspection effort to inspect the top largest modules). While performing as good as ONE in terms of ROI under SNM (since it ranks the smallest modules as the most defective-prone and therefore requires far less code inspection effort than the other models), ManualUp performs the worst among the compared models under all the other situations in Fig. 12 and Table 8.
- *EASC_E, CLA, Bellwether, SC, and FCM perform worse or similar compared with ONE under SNM (PII=20%) and SSC (PCI=20%).* None of them performs better than ONE on any indicator under either SNM (PII=20%) or SSC (PCI=20%), which indicates that they are not comparable with ONE in prediction performance.

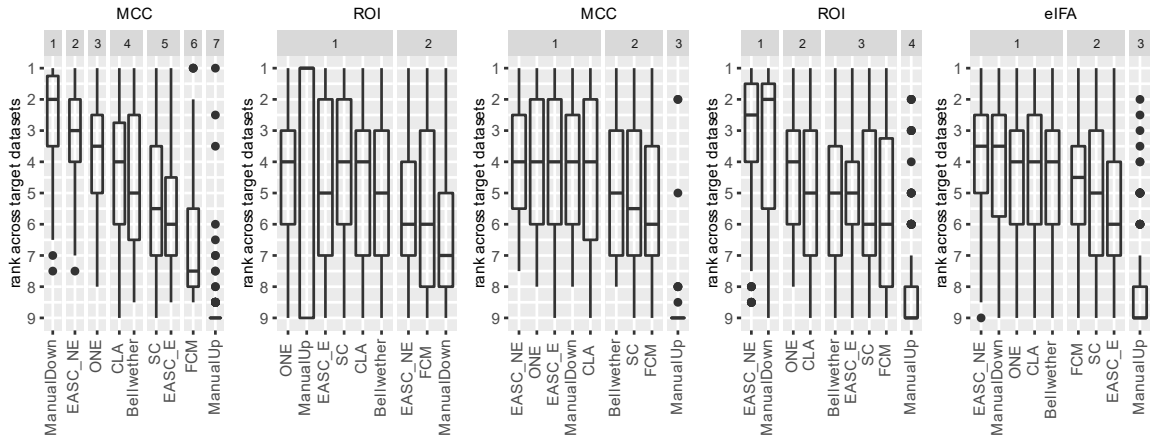

(a) under SNM (PII=20%)          (b) under SSC (PCI=20%)          (c) eIFA

Figure 12: The performance ranking distributions of different models by the non-parametric Scott-Knott ESD test
(The smaller the group number on the top of the plot, the better the performance ranking is)

Table 8: The median value and mean (standard deviation) value of different defect prediction models under MATTER
(The best performance value of each indicator is marked in **bold** while the worst performance value is marked in underlined)

|  | Median | | | | | Mean (standard deviation) | | | | |
|---|---|---|---|---|---|---|---|---|---|---|
|  | MCC (SNM) | ROI (SNM) | MCC (SSC) | ROI (SSC) | eIFA | MCC (SNM) | ROI (SNM) | MCC (SSC) | ROI (SSC) | eIFA |
| Bellwether | 0.178 | 38.6 | 0.108 | 95.6 | 0.002 | 0.175(0.163) | 56.3(59.1) | 0.106(0.117) | 145.3(162.8) | 0.020(0.042) |
| EASC_E | 0.141 | 39.6 | 0.131 | 92.5 | 0.011 | 0.144(0.138) | 73.8(132.3) | 0.122(0.121) | 149.4(170.0) | 0.023(0.042) |
| EASC_NE | 0.251 | 34.7 | **0.151** | **128.9** | **0.000** | 0.253(0.128) | 53.1(53.3) | **0.157(0.116)** | **206.6(215.2)** | 0.019(0.042) |
| SC | 0.129 | 38.2 | 0.081 | 75.3 | 0.007 | 0.144(0.151) | 74.1(141.0) | 0.089(0.119) | 140.0(168.8) | 0.027(0.048) |
| CLA | 0.206 | 34.0 | 0.136 | 122.9 | **0.000** | 0.217(0.133) | 62.2(74.6) | 0.137(0.119) | 160.4(166.3) | **0.017(0.035)** |
| FCM | 0.040 | 31.0 | 0.085 | 56.7 | 0.003 | 0.037(0.177) | 58.6(89.4) | 0.074(0.147) | 132.8(172.8) | 0.028(0.056) |
| ManualDown | **0.268** | 29.0 | 0.142 | 125.5 | **0.000** | **0.284(0.137)** | 47.3(45.6) | 0.141(0.116) | 202.2(218.7) | 0.024(0.048) |
| ManualUp | -0.150 | **102.8** | -0.282 | 17.3 | 0.084 | -0.167(0.097) | **2268.7(11500.2)** | -0.285(0.124) | 52.6(93.5) | 0.118(0.115) |
| ONE | 0.219 | 33.4 | 0.145 | 108.0 | **0.000** | 0.240(0.131) | 57.0(57.7) | 0.150(0.111) | 170.0(187.1) | 0.018(0.039) |

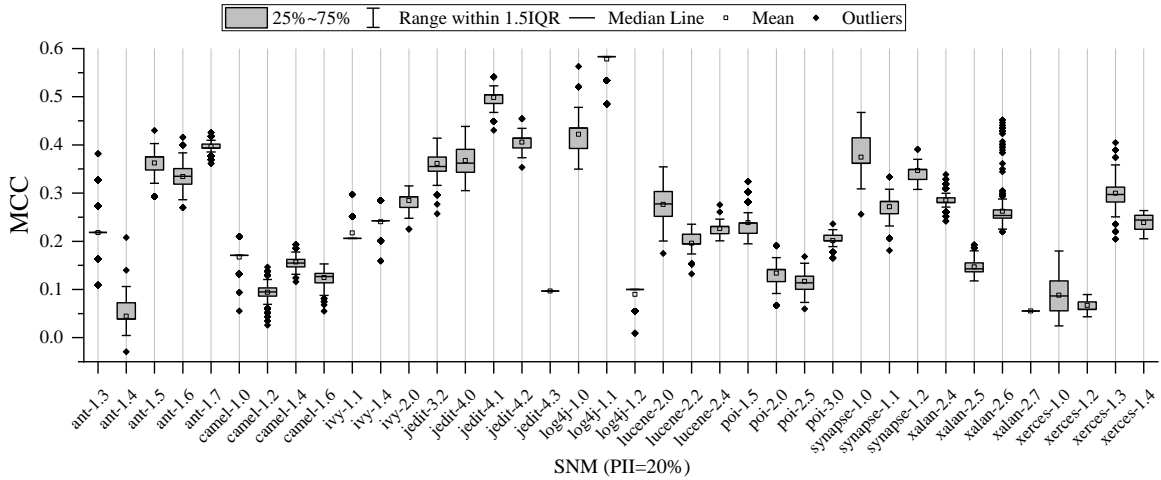

Figure 13: MCC performance distributions of EASE_NE under PII=20% on the test set (i.e., the x-axis) when randomly sampling half sets from all available training sets 500 times in the JURECZKO dataset as the training data

Overall, ONE has a prediction performance strong enough under both SNM and SSC in terms of ROI and eIFA that consider both context switch effort and code inspection effort. Of the compared eight models, EASC_NE and ManualDown are the most promising competitors to ONE. However, they have an inferior performance in terms of ROI under SNM. Furthermore, being a supervised model, EASC_NE has a performance depending on the training data used. In other words, EASC_NE is unable to provide a stable performance for the same test data when the training data changes. As an example, Fig. 13 shows the MCC performance distributions of EASE_NE under PII=20% on the test set when randomly sampling half sets from all available training sets 500 times in the JURECZKO dataset as the training data. As can be seen, the prediction performance of EASC_NE on the same test project varies a lot with the training data used. This pitfall makes it inappropriate to be a global reference point for across-study model performance comparison.

> ***Conclusion.*** *In addition to being "simple" and "stable", ONE is "strong" enough in terms of the defect prediction performance, regardless of whether the SNM or SSC setting is considered. The 3S properties (simple in implementation, strong in prediction ability, and stable in prediction performance) of ONE make it be an appropriate common baseline model used in MATTER.*

### 6.4 RQ4: How far have we really come in defect prediction according to MATTER?

Similar to Fig. 12, Fig. 14 reports the grouped performance ranking distributions of representative defect prediction models and ONE by the non-parametric Scott-Knott ESD test (the smaller the group number on the top of the plot, the better the performance is). Furthermore, Table 9 summarizes the median and mean values of three core indicators under SNM and SSC. The best performance value of each indicator is marked in bold while the worst performance value is marked in underlined. The detailed prediction result for each test set is provided online [94]. The comparison results on individual test data sets can be found in the online Appendix D [83], from which a similar conclusion can be observed with the conclusion drawn on all the datasets.

From Fig.14 and Table 9, we have the following observations.

- *CamargoCruz09-NB, Amasaki15-NB, and Peters15-NB perform similarly to ONE in terms of MCC and ROI but are inferior in terms of eIFA.* Under SNM (PII=20%), those representative models achieve a similar median MCC (range from 0.209 to 0.222 vs. 0.219 for ONE) and a better median ROI (range from 37.2 to 40.0 vs. 33.4 for ONE). However, according to Fig. 14(a), their performance ranking differences are negligible. Under SSC (PCI=20%), those representative models achieve a better median MCC (range from 0.155 to 0.160 vs. 0.145 for ONE) and better median ROI (range from 120.5 to 122.5 vs. 108.0 for ONE). However, according to Fig. 14(b), their performance ranking differences are negligible. In particular, ONE performs significantly better in terms of eIFA.
- *KSETE performs similarly to ONE in terms of ROI under SNM but is inferior to ONE in all other cases.* Under SNM (PII=20%), KSETE achieves a better median ROI than ONE (39.9 vs. 33.4). However, according to Fig. 14(a), their performance ranking differences are negligible. In terms of MCC, ONE performs better than KSETE, i.e., their difference is statistically significant and the effect size is non-negligible. Under SSC (PCI=20%), ONE performs better than KSETE, regardless of whether MCC or ROI is considered. Moreover, in terms of eIFA, ONE also performs significantly better than KSETE.

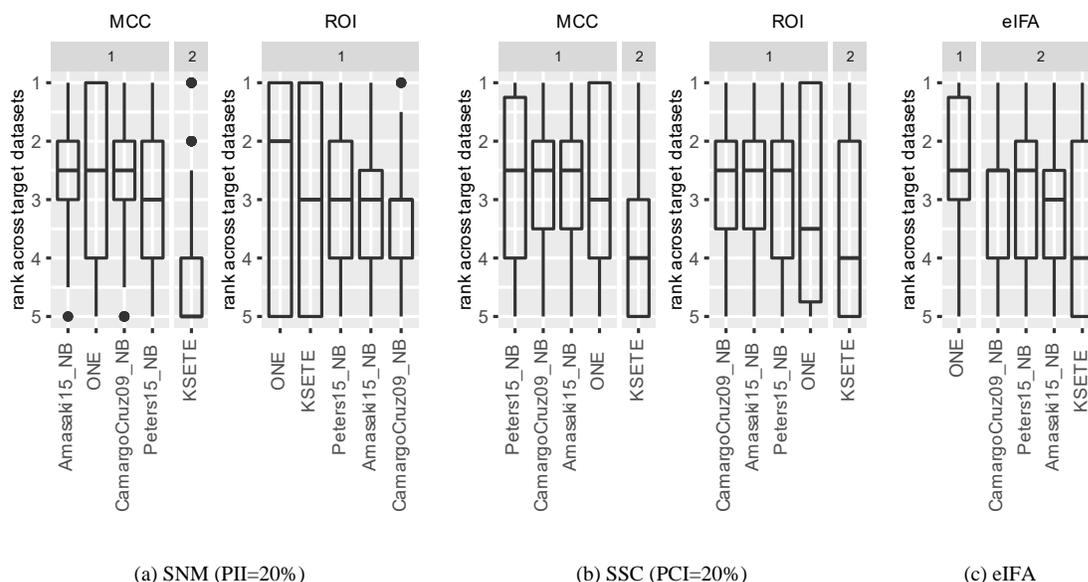

(a) SNM (PII=20%)        (b) SSC (PCI=20%)        (c) eIFA

Figure 14: The performance ranking distributions of ONE, KSETE, CamargoCruz09-NB, Peters15-NB, and Amasaki15-NB by the non-parametric Scott-Knott ESD test (the smaller the group number on the top of the plot, the better the performance ranking is)

Table 9: The median value and mean (standard deviation) value of different models and ONE under MATTER
(The best performance value of each indicator is marked in **bold** while the worst performance value is marked in underlined)

|  | Median | | | | | Mean (standard deviation) | | | | |
|---|---|---|---|---|---|---|---|---|---|---|
|  | MCC (SNM) | ROI (SNM) | MCC (SSC) | ROI (SSC) | eIFA | MCC (SNM) | ROI (SNM) | MCC (SSC) | ROI (SSC) | eIFA |
| KSETE | 0.145 | 39.9 | 0.108 | 97.2 | 0.013 | 0.146(0.093) | 85.1(185.0) | 0.094(0.079) | 160.8(173.3) | 0.019(0.028) |
| CamargoCruz09-NB | 0.217 | 38.2 | 0.157 | 120.5 | 0.009 | 0.229(0.125) | 57.4(63.1) | **0.155(0.111)** | 201.2(215.5) | 0.021(0.033) |
| Amasaki15-NB | **0.222** | 37.2 | 0.155 | 122.0 | 0.011 | 0.233(0.123) | 57.5(62.9) | **0.155(0.104)** | 198.6(207.0) | 0.022(0.032) |
| Peters15-NB | 0.209 | **40.4** | **0.160** | **122.5** | 0.008 | 0.216(0.146) | 59.8(64.4) | 0.154(0.136) | 190.3(198.8) | 0.022(0.041) |
| ONE | 0.219 | 33.4 | 0.145 | 108.0 | **0.000** | **0.240(0.131)** | 57.0(57.7) | 0.150(0.111) | 170.0(187.1) | **0.018(0.039)** |

In our online Appendix B, we use MATTER to evaluate the effectiveness of top-core and MSMDA, another two state-of-the-art prediction models. The experimental results show that top-core and MSMDA do not outperform ONE from the viewpoint of practical application. Combining the above results together, we conclude that the current progress in defect prediction is not being achieved as it might have been envisaged. In other words, the real progress in defect prediction is far from satisfactory. However, all these models are claimed or believed to promote the progress in defect prediction in the literature. If we do not use a consistent framework such as our proposed MATTER to conduct the evaluation, we would not be able to know how far we really have come in defect prediction.

> ***Conclusion.*** *According to MATTER, at least for the datasets we used, none of the investigated models outperforms ONE under SNM and SSC. This means that, if the practical prediction effectiveness is the goal, the real progress in defect prediction is not being achieved as it has been reported in the literature.*

# 7 DISCUSSION

In previous sections, we investigate the effectiveness of SQA-effort alignment approaches, three core indicators, and baseline model ONE in MATTER, and apply MATTER to evaluate representative defect prediction models. In this section, we conduct additional experiments to provide a more comprehensive understanding of the characteristics of MATTER to facilitate researchers and practitioners to improve and utilize it in the future.

## 7.1 How do the parameters affect the effectiveness of MATTER?

According to section 4, there are two important factors in MATTER that may influence the evaluation result, the SQA-effort threshold and the excluded-code-size-percentage of ONE. In this section, we will discuss their influence on the evaluation result of MATTER.

### 7.1.1 The influence of SQA-effort-aligned threshold

Fig. 15 reports how the median MCC and ROI of models change with the increase of the code inspection effort/context switch effort under SNM/SSC. Note that eIFA does not change with the SQA-effort-aligned threshold.

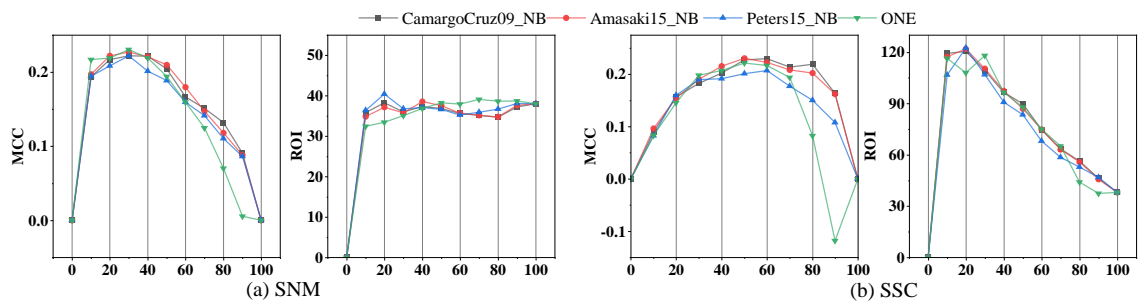

Figure 15: The median performance trend of evaluated models and baseline ONE | varies with the SQA-effort threshold (under SNM and SSC)

Here, we use CamargoCruz09-NB, Amasaki15-NB, and Peters15-NB for comparison, since they show a good performance in RQ4. As can be seen in Fig. 15(a), the comparison results conducted by MATTER are stable when the effort threshold varies from 0% to 50% PCI for CamargoCruz09-NB and Amasaki15-NB, and stable for Peters15-NB when the threshold does not exceed 70%. Similarly, as can be seen in Fig. 15(b), the comparison results conducted by MATTER are stable when the effort threshold varies from 0% to 50% PII for those three supervised models. By varying the effort threshold of MATTER, we reveal that one of the characteristics of these three representative models: at least for the datasets we used, their rankings of the modules in the test projects are not outstanding, especially for the top part of their rankings.

We can see that the comparison results are relatively stable when the available effort is less than 50% PII or 50% PCI. In real software development, it is common that only limited inspection/testing resources are available for SQA. In this case, varying the SQA-effort-aligned threshold in MATTER does not have a large influence on the conclusion.

*7.1.2 The influence of the excluded-code-size-percentage in ONE*

In Section 6, the parameter, excluded-code-size-percentage, in ONE in our experiments is set to 20% in default. In the following, we analyze how the excluded-code-size-percentage affects the performance of ONE and finally the evaluation results of MATTER. Fig. 16 reports how the performance of ONE varyies with the excluded-code-size-percentage under SNM (PII = 20%) and under SSC (PCI = 20%). The median performances are connected by lines. For ease of observation, we mark the median performance value of Peters15-NB, by the red line. The performance of CamargoCruz09-NB, Amasaki15-NB, and Peters15-NB are very close and statistically negligible in our experiment according to RQ4. Therefore we only represent Peters15-NB in Fig. 16 due to the limitation of space.

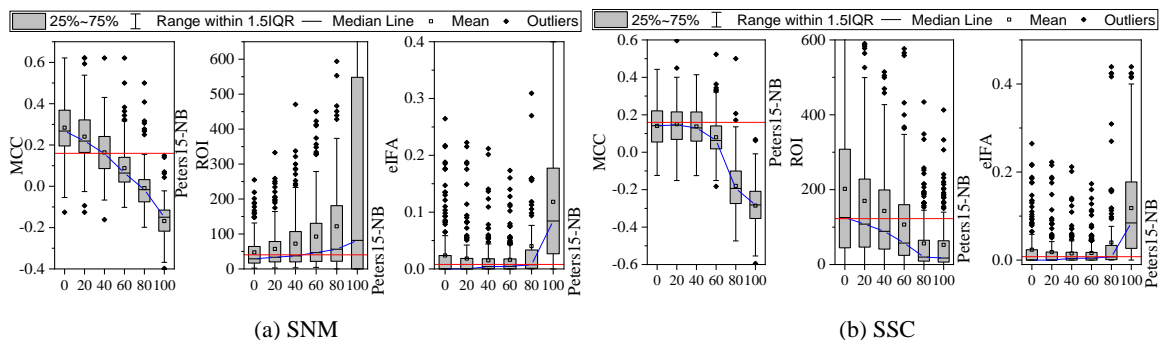

(a) SNM  (b) SSC
Figure 16: The performance trend of ONE with the excluded-code-size-percentage:
(a) under SNM with PII = 20% and (b) under SSC with PCI = 20%

As mentioned before, with the excluded-code-size-percentage getting larger, the size of top modules in N-E is getting smaller. As a result, the total size of the top modules accounting for PII = 20% will become smaller (i.e., PCI will become smaller). Therefore, the recall under SNM tends to decrease with the increase of excluded-code-size-percentage. As shown in Fig. 16(a), the decrease in PCI makes ROI increase with the excluded-code-size-percentage. When the excluded-code-size-percentage grows up and does not exceed 80%, the ROI of ONE first is comparable and then gets better than those three representative models. When the excluded-code-size-percentage does not exceed 20%, ONE performs better than those three models in terms of MCC. In total, under SNM with PII = 20%, for the excluded-code-size-percentage, the default value of 20% leads to a reasonably strong MCC, ROI, and eIFA.

As shown in Fig. 16(b), in terms of MCC, ONE performs best under excluded-code-size-percentage = 20%. After 20%, MCC decreases with the increase of excluded-code-size-percentage. It is clear that the top modules in N-E will become smaller in size when the excluded-code-size-percentage increases. As a result, the top modules accounting for PCI = 20% will contain more modules to be labeled as defective (i.e., PII will become larger). We can see that ROI keeps decreasing due to the increase in PII. Besides, when the excluded-code-size-percentage exceeds 80%, we observe a sharp deterioration of eIFA, indicating that it is hard to find a defective module fast. In total, under SSC with PCI = 20%, for the excluded-code-size-percentage, the default value of 20% leads to a reasonably strong MCC, ROI, and eIFA.

Overall, under both SNM and SSC, for the excluded-code-size-percentage, the default value of 20% SLOC leads to a reasonable performance in terms of effort-aware indicators ROI and eIFA, and the traditional classification indicator MCC when compared with the representative defect prediction models. In other words, our results indicate that 20% SLOC is a appropriate default value to put the top largest modules at the bottom of the module rank in ONE.

### 7.2 How well does ONE perform under the traditional performance evaluation setting?

In RQ4, we find that ONE is a strong baseline under the setting of MATTER. It is natural to ask how well ONE performs under the traditional evaluation settings. To answer this problem, we conduct a comparison of ONE with the representative models under the following setting: (1) use the default classification threshold for each representative model; and (2) use recall, PF, AUC(loc, PD), and AUC(PF, PD). Here, PF denotes the probability of false alarm and PD denotes the recall. AUC(loc, PD) denotes the area under the loc-vs-PD curve (x-axis is the effort measured by the SLOC percentage that the predicted defective modules account for, y-axis is the corresponding PD), while AUC(PF, PD) denotes the area under the PF-vs-PD curve (x-axis is PF, y-axis is the corresponding PD).

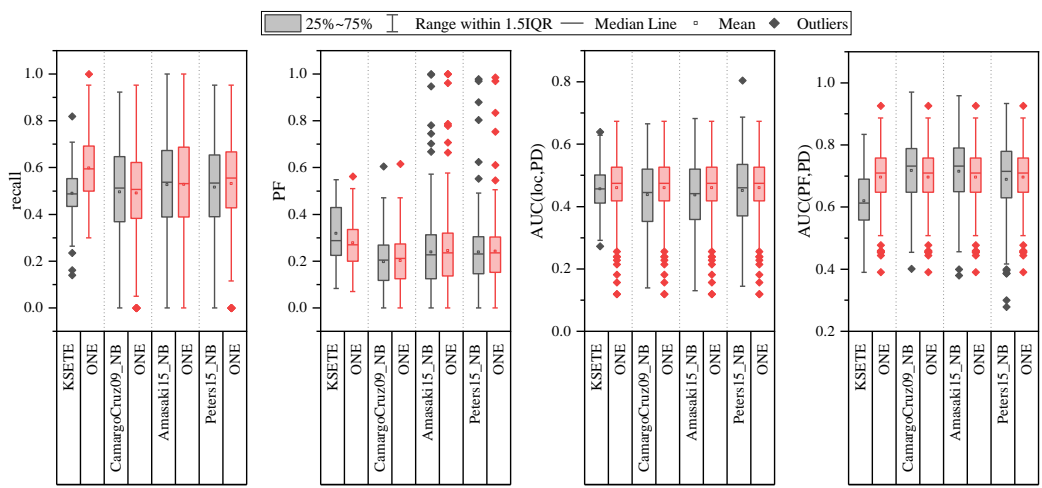

Figure 17: The performance comparison of ONE with the representative models under the traditional performance evaluation setting

Fig. 17 shows the performance distributions of pairs of the evaluated model and ONE aligned by the required PII of the evaluated model in default in terms of recall, PF, AUC(loc, PD), and AUC(PF, PD). Note that we set a classification threshold so that ONE has the same number of predicted defective modules as each compared model. From Fig. 17, we have the following observations. First, ONE has a competitive performance, regardless of which traditional performance indicator is considered. Second, ONE has a median recall around 0.60, a median PF around 0.3, a median AUC(loc, PD) around 0.5, and a median AUC(PF, PD) around 0.7. Note that ONE is an unsupervised model built using only one single metric. In this sense, ONE is a strong baseline, regardless of whether the relative performance or the absolute performance is considered. In summary, ONE is strong enough in defect prediction to provide a "floor performance", helping quickly filter out any model that falls "below the floor".

## 8 IMPLICATIONS

Our study has important implications, as it provides useful guidance on how to carry out an objective evaluation of defect prediction models in practice. The detailed implications are listed as follows:

- **For researchers, our work reveals that there is still a long way to go in defect prediction and MATTER can help guide the healthy and reliable development of defect prediction.**

    Our community constantly updates the state-of-the-art by introducing new defect prediction models. As a result, more than hundreds of defect prediction models have been proposed in the literature. When a new model is proposed, it is common to declared that an advance has been made in the state-of-the-art by showing the superiority in prediction effectiveness over several chosen prior defect prediction models (often recently published). In this sense, substantial progress in defect prediction should have been achieved in the last decades. However, when using MATTER to evaluate the representative defect prediction models, we surprisingly find that most of them are not better than the simple baseline model ONE in terms of the effort-aware prediction performance. This fact reveals that the progress in defect prediction is not being achieved as it might have been envisaged. The fundamental reason for the illusion of real progress, we believe, is the lack of using a consistent way to evaluate the performance of defect prediction models in our community. To tackle this problem, we strongly recommend that our community continuously use MATTER with the same benchmark test data sets in different studies to conduct an inconsistent performance evaluation for any newly proposed defect prediction models. This will certainly help guide the healthy and reliable development of defect prediction.

- **For practitioners, our work reminds them to recognize the value of MATTER in the identification of practical models in the field of defect prediction.**

    Recent years have seen increasing use of increasingly complex modeling techniques for building defect prediction models. For example, much effort has been devoted to introducing deep learning to defect prediction, including deep neural network [105,106], convolutional neural network [32,107,108], and recurrent neural network [109]. The resulting models need to consume a huge computing power and require a large amount of labeled data for training. Undoubtedly, this imposes substantial barriers for practitioners to apply them in practice. This is especially true when considering the fact that those complex defect prediction models often involve a lot of parameters that need to be carefully tuned in practical usage. Our study reminds us that, before applying complex defect prediction models, practitioners could evaluate their practical value via our framework MATTER. These complex defect prediction models are worthy of being applied only when they exhibit considerably better prediction effectiveness than the simple model ONE. Otherwise, the expected return is not sufficient to justify the investment.

- **For tool providers, our work reminds them to integrate the best model identified by MATTER into automatic static analysis tools (SATs) to prioritize static warnings to reduce the high false positive rates.**

    Static analysis tools such as PMD and FindBugs can report code-line-level static warnings without running a program, which have been widely used to detect potential defects in software projects. However, for SQA engineers, when applying SATs to practice, a pain point is that they may produce a lot of false warnings and hence may cause a waste of their precious SQA effort. In [110], Rahman et al. reported that it was feasible to use a module-level defect prediction model to enhance the effectiveness of SATs. Specifically, they ordered the warned lines produced by the SATs with the probability of the modules being defect-prone predicted by a defect prediction model that those warned

lines belong to. Currently, due to the fact that ONE exhibits a remarkable performance, it is promising for tool providers to integrate ONE into SATs to prioritize static warnings to reduce the high false positive rates. Since ONE is a simple module size based model, integrating ONE into the SAT tools would not significantly increase the development cost of the tool providers or the computational overhead of the tool users when using the tool. In the future, if more superior models are developed, we suggest integrating the best model (superior to ONE and simple enough) identified by MATTER into the SAT tools.

In summary, we contribute a framework MATTER toward a consistent performance evaluation in defect prediction. In the future, if any new defect prediction model is proposed, it should be evaluated using our framework MATTER with the same benchmark test data sets to demonstrate its practical value. This will help our community develop practically effective models for predicting defects in a software project and hence really advance the state-of-the-art.

## 9 RELATED WORK

In this section, we introduce the most related works in defect prediction, including the endeavor in the baseline models, model building frameworks, and performance benchmarking.

### 9.1 Baseline models in defect prediction

Recent years have seen several works proposing or investigating the baseline models in the field of defect prediction. In [9], Zhou et al. evaluated the real progress of CPDP by using simple module size based models ManualDown and ManualUp as the baseline models for CPDP models to compare with. They found that "the simple module size models have a prediction performance comparable or even superior to most of the existing CPDP models in the literature, including many newly proposed models." As a result, they advocated that "future studies should include ManualDown/ManualUp as the baseline models for comparison when developing new CPDP models". Specifically, ManualDown is suggested as the baseline model under the non-effort-aware evaluation, while ManualUp is suggested as the baseline model under the effort-aware evaluation. In other words, they indeed suggested two baseline models, separately for non-effort-aware evaluation and effort-aware evaluation. In our study, we suggest only one baseline, ONE, regardless of whether the non-effort-aware or effort-aware evaluation is considered. For practitioners, under ONE, only one single baseline module ranking is needed to consider, which is more coincident with the real development scenario.

Krishna et al. [11] proposed the Bellwether approach for transfer learning and recommended it to be the simple baseline in the field of defect prediction. However, Bellwether is a supervised method whose prediction results depend on the training data set used (see Section 5.2). Therefore, every time the selected "bellwether" project to be used as the training set changes, the prediction of Bellwether will change. As a result, for a given target project, Bellwether is unable to provide a stable prediction. This means that Bellwether is unable to provide "the global reference point" for defect prediction models to compare with. Ni et al. [46] proposed EASC as a baseline model for both effort-aware and non-effort-aware evaluation.

However, when looking at the details, we can find that EASC indeed consists of two models: EASC_E and EASC_NE. The former is suggested as the baseline for effort-aware evaluation, while the latter is suggested as the baseline for non-effort-aware evaluation. In other words, they indeed suggested two baseline models, separately for effort-aware evaluation and non-effort-aware evaluation. For a given target project, EASC provides different module ranks for different evaluation indicators, which is not consistent with the application of defect prediction models in the real world. Moreover, EASC_E and EASC_NE are supervised models and hence do not have the property "stable in prediction performance". Therefore, they are not suitable as "the global reference point" for model evaluation.

## 9.2 Prediction model building frameworks

Song et al. [12] proposed a GPF (General software defect-proneness Prediction Framework). The goal of GPF is to develop a defect prediction model that achieves a high generalization ability by selecting a learning scheme (a combination of a data preprocessor, an attribute selector, and a learning algorithm) with the historical data. Besides, Jing et al. [111] proposed the ISDA/SSTCA+ISDA framework for class-imbalance problems under both with-project defect prediction and cross-project defect prediction. The goal of ISDA/SSTCA+ISDA framework is to develop a defect prediction model that achieves a high prediction performance by dealing with the class-imbalance problem in the training data. However, the goal of MATTER is to make model performance directly comparable across different studies, thus enabling an objective understanding of the real progress in defect prediction our community made. In summary, the main difference between our MATTER framework and their frameworks is: GPF and ISDA/SSTCA+ISDA are used for model building (at the model building stage), while MATTER is used for model comparisons in prediction performance (at the model evaluation/application stage).

## 9.3 Prediction performance benchmarking

In the last decades, a large number of defect prediction models have been proposed, much effort has been devoted to benchmarking the performance of defect prediction models. Lessmann et al. [112] conducted a benchmarking (we name it as NASA-benchmark) on the NASA dataset to investigate the competitive performance of 22 classifiers. Their results showed that the performance of prediction models was not significantly influenced by the classifiers used. Later, Ghotra et al. [113] revisited their work using the cleaned NASA dataset and the PROMISE data set (we name it as cleanNASA&PROMISE-benchmark). Contrary to the earlier finding, Ghotra et al.'s work suggested that some classifiers tended to outperform others. More recently, Herbold et al. [88] conducted a large-scale comparative study to benchmark CPDP models and they named their benchmark kit "Crosspare". Specifically, they replicated 24 CPDP approaches between 2008 and 2015 and conducted a comparison experiment using five different datasets. As a result, they reported that CamargoCruz09-NB, Amasaki15-NB, and Peters15-NB were the best-performing models. Xu et al. [114] conducted a detailed analysis of SDP studies involving clustering methods (we name it Clustering-models-benchmark). Specifically, they applied 40 clustering models to 27 project versions with 3 types of features and used both traditional and effort-aware indicators for performance evaluation. However, none of the above benchmarking works were conducted under an SQA-effort-aligned framework like MATTER. Although those benchmarks reported the performance ranks for many defect prediction models, the results may be inaccurate due to the fact that the comparisons were conducted under the SQA-effort-unalignment threshold setting. Furthermore, it is still unknown the real progress in defect prediction due to the lack of "the global reference point" in the comparison.

## 10 THREATS TO VALIDITY

In this section, we consider the most important threats to the construct, internal, and external validity of our study. Construct validity is the degree to which the independent and dependent variables accurately measure the concept they purport to measure. Internal validity is the degree to which conclusions can be drawn about the causal effect of independent variables on the dependent variable. External validity is the degree to which the results of the research can be generalized to the population under study and other research settings.

## 10.1 Construct validity

The first threat is that our measurement of SQA-effort in MATTER may not actually reflect the required effort in SQA. In MATTER, we use PCI and PII to measure the code inspection effort and context switch effort, respectively. However, the actual SQA process is complicated and the effort spend on inspected/tested modules is difficult to be measured. Modules with the same PCI (PII) maybe not have the same code inspection (context switch) effort in the real world. Moreover, code inspection effort and context effort may not be sufficient to cover the entire effort required in the SQA process. However, our framework of aligning effort-threshold under comparisons still be a groundbreaking exploration and its measurement of code inspection effort and context switch effort is reasonable (though not perfect).

The second threat is that the performance indicators used in MATTER may not successfully correspond to the practical effectiveness of the evaluated defect prediction models. We recommend using MCC, ROI, and eIFA as the core performance indicators in MATTER. Besides, we report traditional performance indicators including PF, recall, F1, and AUC in discussion and in the MATTER replication kit [94]. We believe that the combinations of our effort-aware indicators ROI and eIFA and other traditional indicators correspond to the practical effectiveness of defect prediction models. Also, the detailed prediction results on each target test set of all models in our experiments are shared in our replication kit [94] in case that other performance indicators will be calculated or replication experiments will be conducted in the future. In particular, we encourage other researchers to do the same.

The third threat is the defective labels in the datasets we use may not correspond to the actual defectiveness of modules. In our data sets, the defect labels are mainly obtained by mining software repositories. In practice, the data in software repositories may be incomplete. As a result, there may exist incorrect defect labels in our data sets. Indeed, this problem is inherent to all the studies that use these data sets, not unique to us. Nonetheless, there is a need to use more accurate data sets to mitigate this threat in the future.

## 10.2 Internal validity

The first threat is from the selection bias of the compared defect prediction models, i.e., our selection of baseline defect prediction models for RQ1-RQ3 and our selection of state-of-the-art defect prediction models for RQ4 may not be representative. For RQ1 to RQ3, we choose not only the recommended supervised baseline models in the literature but also those unsupervised models having the potential to be good baseline models. For RQ4, we choose not only the best-performing defect prediction models reported in a recent comparative study but also a state-of-the-art defect prediction model published in very recent years. The above measures ensure that we choose the most representative defect prediction models to investigate RQ1 to RQ4. In this sense, this threat has been minimized as much as we can. Those models are from top journals (such as TOSEM and TSE) and top conferences (such as ICSE and ASE). In this sense, this threat has been minimized as much as we can.

The second threat is the unkown effect of other factors on consistent model performance evaluation. These factors include human aspects (such as confirmation bias, ego, and scientific rivalry) and data aspects (such as poor data quality and poor/incomplete reporting). In practice, it is difficult to avoid these factors completely. For the issue of poor/incomplete reporting, we advocate our community to reported code, datasets, and detailed prediction results on each module of all their evaluated models in their publicly available replication kit just like what we did in this work. For the issues of poor data quality, there have been some works on the influence of data quality on defect prediction and proposal of high-quality datasets [54,93]. Indeed, if our community can assure the ongoing use of MATTER with the same benchmark test data sets in different studies, the above factors will be kept the same in different studies. As a result, their influence will be reduced to a large extent. Nonetheless, there is a need to explore the influence of the above factors on consistent model evaluation, which deserves separate investigations in the future.

## 10.3 External validity

The main threat to external validity is that our findings may not be generalized to other projects. In our study, we conduct our experiments on 179 releases of five defect datasets that contain a diverse range of domains. We conduct the non-parametric Scott-Knott ESD test on the ranking distributions of compared models rather than on the performance indicator values of compared models to reduce the influence of performance values on individual target releases to the comparison results. Moreover, we not only report the experimental results on all datasets as a whole and get meaningful statistical comparison results from a large number of releases but also report the results on individual projects to reduce the influence of datasets on the experimental results. In this sense, we believe that this external threat is within the acceptable range. Nevertheless, we do not announce that our results can be generalized to all projects. To mitigate the threat to external validity, our experiment should be replicated on other defect datasets. We open-source our code and data to make future replication of our study easy [94].

## 11 CONCLUSION AND FUTURE WORK

Defect prediction is a hot research topic, and much effort has been devoted to developing defect prediction models. As a result, more than hundreds of defect prediction models have been proposed, and indeed more and more new defect prediction models are continuously being developed. Undoubtedly, in the positive aspect, this significantly advances our knowledge about how to leverage various information (such as the information from products and the development process) to pinpoint the locations of potential defects in a software project. However, the negative aspect, and also the urgent aspect, is that the current model evaluation is being in a state of chaos: when a new model is proposed, the current practice is to compare it against their preferred baseline models, with their preferred indicators, to demonstrate the effectiveness in defect prediction. Due to the lack of using a widely acceptable baseline model as the global reference point, it is not possible to know the real progress in defect prediction. Furthermore, many models are compared under the SQA-effort-unaligned threshold setting, resulting in an unfair comparison. Consequently, unintentionally misleading conclusions on advancing the state-of-the-art may be drawn, which can cause missed opportunities for real progress in defect prediction.

In order to tackle the above problem, our study proposes a framework MATTER toward a consistent performance comparison for defect prediction models. In this framework, we use an unsupervised model ONE, which is simple to implement, strong in prediction ability, and stable in prediction performance, as the common baseline model for providing the global reference point. Furthermore, we advocate using the SQA-effort-aligned threshold setting and two effort-aware performance indicators that consider both code inspection effort and context switching effort to make a fair and practical performance comparison. In particular, we use MATTER to evaluate the effectiveness of a number of representative defect prediction models, including the best-performing models reported in a recent comparative study and the state-of-the-art models. Surprisingly, we find that, when the SQA effort required is considered, none of these representative models exhibits a remarkable performance than the simple baseline ONE. In fact, many of them have an inferior prediction performance. On the one hand, this informs us that the real progress in defect prediction is not being achieved as it might have been envisaged, i.e., there is still a long way to go in defect prediction. On the other hand, more importantly, this reveals the importance of our framework MATTER in ensuring the healthy development of defect prediction. Last but not least, it must be pointed out that the burden for applying MATTER to practice is very lightweight: (1) the baseline model ONE is cheap to be built based on the test data, independent from the training data; (2) it is easy to apply the SQA-effort-aligned threshold setting; and (3) it is cheap to compute the core performance indicators. As a result, we strongly recommend that, in future studies, when any new defect prediction model is proposed, MATTER should be used to evaluate its actual usefulness in order to promote scientific progress in defect prediction. We make our implementation of MATTER publicly available to support future research.

In the future, we plan to investigate the practical effectiveness of more defect prediction models by MATTER. Our purposes are two folds. First, we aim to help practitioners to find which models are really useful. Second, we aim to help researchers to determine which modeling techniques are promising in defect prediction and hence help them further develop more effective techniques.

## ACKNOWLEDGMENTS

We are very grateful to (1) the authors of Bellwether, SC, CLA, and KSETE for sharing their implementation code; (2) Herbold, Trautsch, and Jens Grabowski for sharing the implementations of CamargoCruz09-NB, Amasaki15-NB, and Peters15-NB; and (3) the authors of FCM and EASC for providing the detail descriptions on their models. In particular, we thank Dr. Ni for illustrating the implementation detail of EASC by communication, ensuring the accurate implementation of EASC in our study. Last but not least, we want to point out that our purpose is not to criticize any existing work but instead to work together as a community to get through the severe and urgent challenges in consistent defect prediction evaluation we face today.